\documentclass[aps,prb,showpacs,raggedbottom,nobalancelastpage,amssymb,twocolumn,longbibliography]{revtex4-1}
\usepackage{amsmath}
\usepackage{amsfonts}
\usepackage{graphicx}
\usepackage{float}
\usepackage{appendix}
\usepackage{dsfont}
\usepackage{amssymb}
\usepackage{bm}
\usepackage{color}
\usepackage{soul}
\usepackage{natbib}
\setstcolor{red}
\usepackage{nccmath}
\usepackage{array}

\newcommand{\beq}{\begin{equation}}
\newcommand{\eneq}{\end{equation}}
\newcommand{\be}{\begin{equation}}
\newcommand{\ee}{\end{equation}}
\newcommand{\bea}{\begin{eqnarray}}
\newcommand{\eea}{\end{eqnarray}}

\usepackage{multirow}
\usepackage{wasysym}


\makeatletter
\def\@bibdataout@aps{%
\immediate\write\@bibdataout{%
@CONTROL{%
apsrev41Control%
\longbibliography@sw{%
    ,author="08",editor="1",pages="1",title="0",year="1"%
    }{%
    ,author="08",editor="1",pages="0",title="",year="1"%
    }%
  }%
}%
\if@filesw \immediate \write \@auxout {\string \citation {apsrev41Control}}\fi 
}
\makeatother

\begin{document}
\title{Analytical and cellular automaton approach  to a generalized SEIR model
  for infection spread in an open crowded space}
\author{Andrea Nava, Alessandro Papa, Marco Rossi, and  Domenico Giuliano}
\affiliation{
Dipartimento di Fisica, Universit\`a della Calabria, Arcavacata di Rende I-87036, Cosenza, Italy \\
INFN - Gruppo collegato di Cosenza,
Arcavacata di Rende I-87036, Cosenza, Italy
 }
\date{\today}

\begin{abstract}
We formulate a generalized susceptible exposed infectious recovered (SEIR) model on a graph, describing   the population dynamics of an open crowded place with an
arbitrary topology. As a sample calculation, we discuss three simple cases, both analytically, and numerically,   by means of a cellular
automata   simulation of the  individual dynamics in the system. As a   result,
we provide the infection ratio in the system  as  a function of controllable parameters,
which allows for quantifying how  acting on the human behavior may effectively   lower the disease spread
throughout the system.
\end{abstract}

\maketitle

\section{Introduction}
\label{intro}

Compartmental models provide a conceptually simple and widely used mean to mathematically  modeling
the dynamics of infection transmission in isolated populations~\cite{SIR0,SIR1,SIR2,SIR3}.
In such models, the population is divided into compartments, each one corresponding to a specific
health status of the individuals that belong to it. For instance, the basic SIR model consists of
three compartments,  the  {\it Susceptible}  compartment ($S$), to which
not infected, healthy individuals belong, the
{\it Infectious}  compartment ($I$), that contains infectious individuals,  and  the {\it Recovered}  compartment ($R$),
that contains individuals who either recovered from the infection, or
 died.  In the SIR model,
the spread of the infection is described in terms of a set of
differential equations that describe the population transfer  from one compartment to another one. The
key parameters of the model are the  rates of transfer between the  compartments which, in general,
are quantities to be experimentally fitted.

In fact, while the SIR model is a good tool  for long time simulations, it does not take into account  the {\it latent} phase,
in which people are infected but not yet infectious. The simplest way to fix this flaw of the model is by
adding the {\it Exposed}  compartment ($E$). The $E$ compartment  contains individuals who have been infected, but are not yet infectious.
This improvement eventually leads to the   SEIR model, more appropriate for short-time simulations.
For instance, the SEIR model has recently been widely employed to describe the Covid-19 infection, though with
some limitations~\cite{berger,yang,ala,fodor}.

Despite their effectiveness in describing a number of real-life infection dynamics, the SIR and the SEIR models,
together with their generalization to a higher number of different compartments~\cite{heth}, are in general
affected by a limitation. Indeed, they describe, in general, the average  global behavior of the population,
with no attention to local dynamics that can make the level of infection strongly depend on the region in
real space that one is considering. Instead, this last aspect has become of fundamental importance in, {\it e.g.},
countries recently affected by Covid-19 pandemic, such as Italy, which show a strongly uneven spread
of the infection across the country territory [see, for instance,  https://en.wikipedia.org/wiki/COVID-19\_pandemic\_in\_Italy
for a detailed description of the Covid-19 infection distribution across Italy from the start of March, 2020].
In addition, the SIR- and the SEIR-type models basically describe closed systems,
without the possibility for people to enter or exit from the system. So, they are also expected not to be
reliable when describing relatively small populations that can exchange individuals with the surrounding
``environment'', such as shopping centers or closed commercial areas. In systems as such,  the infection dynamics is again
expected to be strongly space-dependent, due to, {\it e.g.}, the presence of very popular shops in a shopping mall, where
people spend on the average much more time than in different areas.

In general, it is well known that the spread of a disease has a strong dependence upon
the topology of the system and on how individuals and/or  subgroups are connected to each other.
Specifically, the topology strongly affects the rates that eventually enter the differential
equations describing the population dynamics and, therefore, it determines the specific stationary solution,
describing the system over long time scales~\cite{small}. For this reason, in the last years, a remarkable
amount of work has been devoted to discussing the dynamics of epidemic processes in metapopulation models \cite{meta0, Soriano, meta1, meta2,meta3,meta4} on graphs
and hypergraphs~\cite{network11,network10,network6,network2}. 
However, previous metapopulation   movement-contagion descriptions focus on large-scale models, rather than on local 
traffic flows, that are fundamental for the description of the infection spread in small areas.
In these systems,
time- and space-inhomogeneity  leads to nontrivial consequences on the
population dynamics~\cite{network7,network9,network8,network5,network3,network1}. These are
of the utmost importance over  short-time,  small-space scales, as in the case of the  disease dynamics on a sidewalk,
in a shopping center or running tracks, where the topology of the
system strongly affects the effective contact rate, that is, the number of contacts, per unit of time,
between individuals~\cite{one-way}.

Based on the above observations, in this paper  we describe the population dynamics of an open crowded place
in terms of a ``local'' SEIR model on a graph, with position-dependent parameters.  In particular, after writing  the set of differential equations describing
our system, we first provide an approximate analytical solution within a mean field approach to
the full mathematical problem and, therefore, we resort  to a cellular automata (CA) simulation of the people traffic in the system~\cite{CA1,CA2,CA3}.
 In general,  an analytical treatment of the problem in terms of a collection of local
compartmental models in a nonuniform nonequilibrium configuration is quite difficult to deal with, as the number of differential
equations scales quickly with the size of the system. At variance,  the CA is able to describe the pretty complicated
behavior in terms of simple rules, that apply simultaneous to all the nodes of the graph.
Eventually, we employ the CA numerical results to pertinently complement and check
the reliability of the (approximate) analytical ones.

To quantitatively describe the infection spread over the graph, in all the cases
we compute the (average) infection ratio in the steady state of the system,
$C_r$, that is, the ratio between the variation in the average population in the
compartment $E$ over a  time $T$ long enough for the system to reach the steady state and
the initial average population in the compartment $S$.  $C_r$ measures the hazard for a healthy
individual to get infected when going across the system in the
presence of infected individuals. Therefore, to quantify the level of infection risk
for an individual in the system, we compute $C_r$ as a function of various system parameters,
some of which are particularly important, as they can be readily acted on by, for instance,
controlling the entrance rate of people in a shopping center, letting people move in one direction
only at each side of the aisles in the center,  increasing or decreasing the number
of points at which it is possible for individuals to change their direction of motion, and so on.

The key feature of our work is that here we focus onto open graphs. 
Differently from, e.g., the models studied in Refs.[\onlinecite{meta0, meta1, meta2,meta3,meta4}] and in 
Refs.[\onlinecite{network11,network10,network6,network2}], in our system,  open boundary conditions 
 introduce a characteristic timescale, corresponding to  the average time  that people spend inside the 
 system. As we show throughout our derivation, this time has  nontrivial effects on   the diffusion of the  infection. When possible, 
 we analytically compute it,  otherwise we resort to a numerical calculation so to show how to act on it 
to reduce the infection spread. Eventually, we  suggest how to apply metapopulation 
models to simulate a small open crowded space, proposing how to relate transition 
probability to experimentally measurable and adjustable quantities. Doing so, we provide the possibility to experimentally 
verify our claims in real life situations and act according, to reduce infection spreads.

Considering all together in our model the effects of both   effective parameters that depend on
the (``intrinsic'') disease dynamics and of parameters that can be tuned by acting on the social behavior of
individuals, we are able to quantify how accurate control of
human-dependent effects, such as  social distancing, use of personal protective equipment and so on, can be effective in
mitigating the effects of high contagion rates of diseases, such as the one due to Covid-19~\cite{Covid1,Covid2,Covid3,Covid4}.

To present the main features of our model and to
discuss its implementation, both analytical and numerical, in the paper we focus on three
simple, prototypical models of open systems. Yet, as we eventually discuss in the paper,
within the CA approach, generalizations of our models to more realistic situations
are straightforward, and we plan to  pursue them in forthcoming publications.
In the paper we use the expression ``cellular automata'' in the same sense it is typically used in traffic flow models  \cite{cellulartraffic}: 
all pedestrians are indistinguishable and the evolution rules do not depend on the “history” of 
each individual but only on the cell they occupy and and on the state of the  neighboring ones. However, 
in each cell we have not a Boolean variable $n_i$ (occupied/unoccupied cell) but a (discrete) multi-value density 
distribution function, that corresponds to a coarse-graining description of the pedestrian crowd. In this sense 
our approach is more similar to a lattice Boltzmann model that, historically, was developed as an extension of the CA 
model and, at the same time, represents a discretization of the classical continuous Boltzmann equation \cite{CALB, latticeboltzmann}.

The paper is organized as follows:

\begin{itemize}
 \item In Sec.~\ref{theo},  we define our generalized SEIR model on graphs.  In particular, we
present the (local) set of differential equations describing the (local) population dynamics on
the graph and provide an explicit analytical mean-field solution of the equations. Eventually,
after presenting the corresponding results, we highlight the main limitations of the analytical
approach, which motivate our switch to the numerical, CA approach;

\item In Sec.~\ref{rule}, we present and discuss the CA rules  describing the spread of an infection within an open,
finite connected graph. Therefore, we implement them to obtain numerical results in the same systems  analytically studied in Sec.~\ref{theo}. Eventually, we employ the numerical results to check the reliability
of the analytical approach we use in  Sec.~\ref{theo};

\item In Sec.~\ref{conclusion}, we provide our main conclusions and discuss about possible further perspectives
of our work;

\item In Appendix  \ref{rseir} we review the basic formulations of the  compartmental SEIR model.

\end{itemize}

\section{The lattice local SEIR model}
\label{theo}

In this section, we define and analytically study  our lattice model generalization of the  SEIR model,
suitable to   describe the spread of an infection throughout a number of open, finite connected graphs.
Within our model, we  describe the dynamics of small populations, each one residing at the sites of a pertinently
designed (quasi)one-dimensional lattice, and connected to each other by means of a finite
rate for individuals to ``hop'' from one site to the others. As we discuss in the following,
our model is able to encompass several typical features of
infection spread in real world, particularly evidencing, on quantitative grounds, how
the spreading depends on in principle tunable parameters of the system.

Throughout this paper, we focus onto linear graphs, that is, finite one-dimensional lattices (chains),
with open boundary conditions. Each chain corresponds to a simplified model of
a straight way for pedestrians. More complex (and, in many cases, more realistic) models
can be readily constructed by, {\it e.g.}, putting together finite, one-dimensional lattices to be
used (with the appropriate boundary conditions) as ``elementary building blocks''.
Each lattice site $j$ hosts a ``local'' population of individuals that is characterized by the
density of healthy people $n_{S , \lambda ; j}$, that is, the number of healthy people per plaquette, by the density of exposed (infected, not contagious)
people  $n_{E , \lambda ; j}$, that is, the number of exposed individuals per plaquette,  and by the density of infectious people $n_{I , \lambda ; j}$,
that is, the number of infectious individuals per plaquette. In this respect, each
cell realizes a population whose dynamics is described by the SEIR model reviewed in appendix~\ref{rseir}, though with a
major simplification which we discuss next. In fact,  the basic
assumption of the SEIR model that only an infectious individual can turn into a
recovered one (either healed, or deceased), as well as with the observation that
the duration of the time that each individual spends within the graph is of a few hours, which is pretty short
compared with the time scales for having a nonzero density of recovered people,
enables us to set to  zero the density of recovered people throughout the whole
system. Furthermore, as the median incubation period of a virus such as 2019-nCoV ARD has been estimated
to be about 3.0 days~\cite{Covid5}, we set to zero the probability for exposed people to become infectious.
Accordingly, we assume that the total number of infectious individuals in
the system can only change by exchanges with the outside (infectious individuals
either entering, or exiting, the system).

Compared to the ``standard'' SEIR model, our framework allows for the net number of individuals in
the population at each lattice site to change, as a consequence of individuals hopping between neighboring
sites on the chain. To formalize this aspect of our system, when defining the various local densities, we add a label
$_\lambda$, encoding various possible ways for individuals to move between different sites.
In general,  in our sample models, at any time $t$, each individual
at site $j$ of a chain can either move towards the right (corresponding to $\lambda > 0 $),
  or towards the left (corresponding to $\lambda < 0$),  on one of the possible parallel
  lanes labeled by $\left|\lambda\right|=1,...,\Lambda$,  with
rules and constraints that depend on the specific lattice topology.

Regardless of the specific lattice topology, the mathematical description of the population dynamics on
the lattice can be formalized by a set of differential equations, whose parameters are determined as
follows. First of all, to ease the mathematical formulation, it is useful to label each cell by
both the lattice site index, $j$, as well as with an additional index $\lambda$ encoding
 the information about the direction of motion. So, each ``physically
distinct'' lattice site $j$ corresponds to  several cells, which we label with the pair of indices
$j , \lambda$. Within each cell,
$n_{S , \lambda ; j}$, $n_{E , \lambda ; j}$, and $n_{I , \lambda ; j}$ can either be zero, or
different from zero.

\begin{figure}
\center
\includegraphics*[width=0.8 \linewidth]{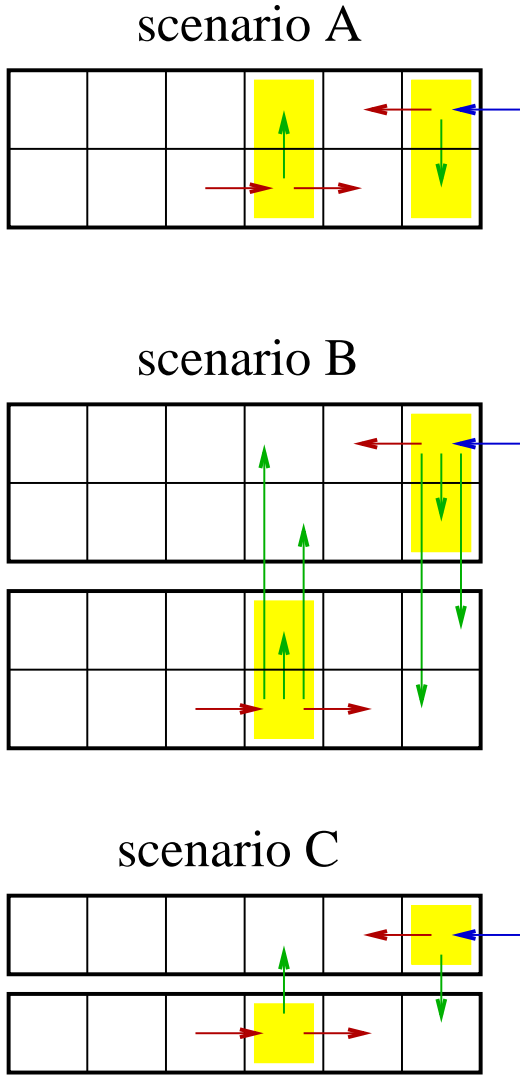}
\caption{Scenario A: people moving over a narrow street or a sidewalk;
they can walk in both directions and are close together at a distance
less than the safety distance $d$. The arrows pictorially encode the cellular automata rules of Sec.~\ref{rule} that act every time step $\Delta t$.
In particular, the top row corresponds to people moving to the left, the
bottom row to people moving to the right.
In the bulk of the system, individuals can make a step  to the next cell
with probability $p_h$ (red arrow), or change their direction, with probability $p_{\ell}$  (green arrow). At the boundaries, an individual
can enter into the system with a probability $p_{\rm in}$ (blue arrow). During the contagion phase, an healthy individual
can be infected by infectious individuals moving in either direction  (here, as well as in the scenario B and in scenario C,
the yellow background represents the range of the infection). Scenario B is  like scenario A,  but with people
moving in both directions  distributed along the two sides of the street (the ``sublattices''). People can change their
direction of motion or change street side or both (the probabilities, $p_{\ell}$, associated to the three processes are, in principle, different) but,
if they lie on different sublattices, they
 cannot infect each other.
Scenario C is like scenario B, but each sublattice hosts individuals moving in one direction only. The cellular automata
probabilities are related to the rates of the theoretical model of Sec.~\ref{theo} by $p_A = \omega_A \Delta t$.}
\label{CA}
\end{figure}
\noindent
At any cell  $j , \lambda$ not residing at the endpoints of the chain (that is,
with $j \neq 1,L$),  on top of the ``standard'' SEIR dynamics for an isolated
population, the number of individuals in each compartment can either change because
individuals hop  from-, and into-,
neighboring cells {\it moving in the direction defined by  $\lambda$},
or {\it by changing the sign of $\lambda$} (that is, the direction of motion) at a given $j$.
To formally describe individual motion between different cells, we set $\omega_{h;\lambda}$ to be the rate (probability per unit time) of an individual to hop from
cell-$j,\lambda$ to cell-$j\pm 1, \lambda$ (where sign plus is for $\lambda>0$ and minus for $\lambda<0$)
and $\omega_{\ell;\left( \lambda,\bar{\lambda} \right)}$ to be the rate for an individual
to hop from cell-$j ,\lambda$ to cell-$j,\bar{\lambda}$, with $\bar{\lambda} \neq  \lambda$.
If $\lambda \bar{\lambda}>0$, the individual is changing its walking lane but not its direction of motion,
while, if $\lambda \bar{\lambda}<0$ the individual is changing both lane and direction of motion.
To simplify our further derivation, we assume that the system is homogeneous in real space,
which implies that the  rates are all independent of the index $j$. Moreover, as there
is apparently no reason for different types of individuals to move at different rates,
we assume that $\omega_{h;\lambda}$ and $\omega_{\ell;\left( \lambda,\bar{\lambda} \right)}$
are  independent of whether the moving individual is $S$, $E$, or $I$.

As we aim at eventually describing steady states of the
system, without big local fluctuations in the various densities, consistently with the  detailed balance
principle, we  assume $\omega_{\ell;\left( \lambda,\bar{\lambda} \right)} =
\omega_{\ell;\left( \bar{\lambda} , \lambda \right)}$. Finally, to account for the
``local'' SEIR dynamics, we introduce the parameter $\omega_c$, which corresponds to
the infection rate that determines the change in time of $n_{S , \lambda ; j}$, $n_{E , \lambda ; j}$, and $n_{I , \lambda ; j}$
at given $j , \lambda$. At the endpoints of the chains, that is, for  $j=1$, or $j=L$,   the rate for individuals (of any type) to enter the cell
at fixed $\lambda>0$ ($\lambda<0$) is simply the entrance rate into the system, $\omega_{\rm in}$ (which is one of the tunable parameters of
our system). Similarly, for  $j=L$ ($j=1$), the rate for individuals (of any type) to exit the cell
at fixed $\lambda>0$ ($\lambda<0$) is given by  the exit rate from the system, $\omega_{\rm out}$.
The rules listed above allow us to fully determine the set of equations describing our model. In addition, they also
completely define the CA rules, once the rates are traded for the corresponding probabilities at each elementary
time step, by multiplying all of them by the elementary time step of the CA, $\Delta t$ (see Sec.~\ref{rule} for an
extensive discussion of this point). Therefore, for the sake of our presentation, we pictorially present all the
above rules in  Fig.~\ref{CA} which, as stated above, applies to both the mathematical model and to
the CA.

Putting the various ingredients listed above all together, we obtain the
following set of differential equations  for the local densities on the lattice, for individuals
moving in both directions and for $1<j<L$:

\begin{eqnarray}
\frac{dn_{S,\lambda;j}}{dt} & = & \omega_{h;\lambda}n_{S,\lambda; [ j - {\rm sgn} ( \lambda) ]  }
-\omega_{h;\lambda}n_{S,\lambda;j} \nonumber \\  
& &+ \sum_{\bar{\lambda} \neq \lambda}\omega_{\ell;\left( \lambda,\bar{\lambda} \right) } ( n_{S,\bar{\lambda};j}- n_{S,\lambda;j})  -\omega_{c}
f_{\lambda;j}\left(\left\{ n_{\nu,\lambda;j}\right\} \right) ,  \nonumber \\
& & \label{list.1} \\
\frac{dn_{I,\lambda;j}}{dt} & = & \omega_{h;\lambda}n_{I,\lambda; [ j - {\rm sgn} ( \lambda) ] }-\omega_{h;\lambda}
n_{I,\lambda;j} \nonumber \\
& & + \sum_{\bar{\lambda} \neq \lambda}
\omega_{\ell;\left( \lambda,\bar{\lambda} \right) } ( n_{I,\bar{\lambda};j}- n_{I,\lambda;j}) \:\: , \label{list.2}  \\
\frac{dn_{E,\lambda;j}}{dt} & = & \omega_{h;\lambda}n_{E,\lambda;[ j - {\rm sgn} ( \lambda) ]  }
-\omega_{h;\lambda}n_{E,\lambda;j} \nonumber \\
& &+\sum_{\bar{\lambda} \neq \lambda}
\omega_{\ell;\left( \lambda,\bar{\lambda} \right) } ( n_{E,\bar{\lambda};j}- n_{E,\lambda;j}) +\omega_{c}f_{\lambda;j}\left(\left\{ n_{\nu,\lambda;j}\right\} \right)
 . \nonumber \\
& & \label{list.3}
\end{eqnarray}

\noindent
Consistent with the rules we discuss above, for  $j=1$ ($j=L$) and $\lambda>0$ ($\lambda<0$), the terms
$\omega_{h;\lambda}n_{\{S , I , E\} ,\lambda; [ j - {\rm sgn} ( \lambda) ]  }$ at the
right hand side of Eqs.~(\ref{list.1})-(\ref{list.3}) must be
replaced with $\omega_{{\rm in};\{S , I , E \} ,\lambda}$. Similarly,  for
$j=L$ ($j=1$) and $\lambda>0$ ($\lambda<0$), the terms
$-\omega_{h;\lambda}n_{\{S , I , E \} ,\lambda;j}$ at the
right hand side of Eqs.~(\ref{list.1})-(\ref{list.3}) must be
replaced with $-\omega_{{\rm out};\lambda}n_{\{S , I , E \} ,\lambda;L(1)}$.

A general comment about the set of Eqs.~(\ref{list.1})-(\ref{list.3}) is that, while they
certainly apply for low values of the individual densities at each cell, it is
reasonable to assume that the {\it maximum total density} of individual at each cell does not
go beyond a maximum value $n_{\rm{max}}$ (which, by symmetry, we assume to be cell-independent), that
is, $n_{S , \lambda ; j} + n_{E , \lambda ; j} + n_{I , \lambda ; j} \leq n_{\rm{max}}$. Formally, such a
constraint can be easily implemented in the set of differential equations above by, {\it e.g.}, substituting
$n_{\{ S , I , E \} ,\lambda;j}$ at the right-hand side of the equations  with $n_{\{ S , I , E \} ,\lambda;j} ( n_{\rm{max}} - \sum_{B} n_{B,\lambda;j \pm 1} )/ n_{\rm{max}} $.
In fact, while we definitely take into account the constraint when solving the equations, as well as when defining
the cellular automata rules below, to ease the notation we prefer not to explicitly write it down in Eqs.~(\ref{list.1})-(\ref{list.3}),
by limiting that set of equations to the low-density regime.
It is worth stressing that, with $n_{max}$ equal to $15$, saturation effects hardly emerge  
in the low-density high-mobility regime ($\omega_h>0.05$), in any cell of the system. On the other side, at 
small values of $\omega_h$  ($\omega_h<0.05$) the system easily saturates in the entrance cell, for any value 
of $\omega_{in}$.  In this case, if an infectious individual is present, the contagion can spread to all the other people.
Clearly, the low-density regime is more realistic and accurate. To avoid unrealistic saturation effects at small
$\omega_h$ it is possible to effectively correct $\omega_h$ as a function of the cell occupation level. Indeed, 
it is more realistic to assume that, if a cell is crowded, real people prefer in general to move forward rather than
creating a traffic jam. This can be easily implemented by replacing the fixed value of $\omega_h$ with a higher effective 
value when $n_i \lessapprox n_{max}$, together with a change in the direction of motion, so that people prefer to move
toward the nearest exit rather than venturing into the crowd. Yet, this does not affect the final  result, when
the saturation is not reached, and has accordingly a negligible effect in the high-density regime.
The function $f_{\lambda;j}$ in Eqs.~(\ref{list.1}), (\ref{list.3})  is the joint probability to truly have infectious and healthy people at the same time in the
same cell. In general, this function is not simple to derive, especially because it changes from scenario to scenario, that is, since it
is strongly affected by the specific details of the lattice. However, at least in the sample cases we discuss below, we show that
 $f_{\lambda;j}$ can be effectively estimated by means of a simple, mean-field approximation.
To be specific, we now present and discuss the  form of the above rate equations in
the sample cases we deal with in our work. In particular,  in the following we consider three different scenarios, that is

 \begin{itemize}
  \item {\it Scenario A:} this describes  a small sidewalk with individuals that can move both ways within one chain only;

  \item {\it Scenario B:} this describes, for instance, a wide shopping mall, or an aisle in a shopping center. In this case,
  individuals are more or less evenly distributed  between the two sides of the street, in front of  the shop windows, so that people on different
sides cannot infect each other;

\item {\it Scenario C:} this describes a couple of reverse one-way streets. Basically, in this
case   people with different walking directions are forced  to stay on different
sides of the street at a distance greater than the safety distance, so that people moving toward opposite directions
cannot infect each other.
   \end{itemize}
Depending on  the specific scenario we are focusing on,
we resort to different mean-field decouplings for
$f_{\lambda;j}\left(\left\{ n_{\nu,\lambda;j}\right\} \right)$. In general,
the mean-field approximation for the joint probability function is grounded on the ansatz

\begin{equation}
f_{\lambda;j}\left(\left\{ n_{\nu,\lambda;j}\right\} \right)\thickapprox n_{S,\lambda;j}\left(n_{I,\lambda;j}+\sum_{\bar{\lambda}\neq \lambda}
\mu_{\left( \lambda,\bar{\lambda} \right)} n_{I,\bar{\lambda};j}\right)
\:\:\:\: ,
\label{truly.2}
\end{equation}
\noindent
with the $\mu_{\left( \lambda,\bar{\lambda} \right)}$'s, that are either equal to
 zero, or to  one, encoding all
the specificities of each case. In particular, case-by-case we
choose the $\mu_{\left( \lambda,\bar{\lambda} \right)}$'s as follows:

\begin{itemize}
\item {\it In scenario A} we have just one  chain on which individuals can move in two opposite directions.
Only one pathway is   available in either direction, so, $\lambda$ can only take the values $\pm 1$ and
$\mu_{\left( \lambda,\bar{\lambda} \right)}=1$.

\item {\it  In scenario B}, in its simplest version, we consider two pathways per each direction of
motion of individuals. Therefore, $\lambda = \pm 1, \pm 2$ and $\mu_{\left( \lambda,\bar{\lambda} \right)}=\delta_{\lambda,-\bar{\lambda}}$.

\item {\it In scenario C} again we have just one  chain on which individuals can move in two opposite directions, such as
in scenario A, but now spatial separation makes it impossible for individuals moving toward opposite directions to
infect each other. Accordingly, here again $\lambda = \pm 1$, but now   $\mu_{\left( \lambda,\bar{\lambda} \right)}=0$.
\end{itemize}

In principle, given the appropriate initial  conditions and once the
various rates have been pertinently estimated, the system of differential equations reported above
is enough to discuss in detail the evolution in time of the individual flows across the system.
In practice, knowing, for instance, the response in time to a sudden change in the system parameters
(the rates) can help to predict the increase/reduction in the infection diffusion once local boundaries between
regions in the same country, or between different countries, are relaxed/enforced (that has recently become
an ubiquitous procedure to  keep the infection under control all around the world). While we
plan to discuss these features in a forthcoming publication, here we are mostly focused onto
infection propagation in an environment where people flow is expected to shortly become
stationary in time. For this reason, here we just focus onto stationary solutions of Eqs.~(\ref{list.1})-(\ref{list.3}).
In fact, while  possible nonuniformities in the stationary density distribution due to
boundary effects (which are in any case negligible in the large system limit) can arise, we numerically checked that at the
system boundaries, $j=1,L$, the asymptotic values of the population densities are the same as in the bulk of the system,
that is, for  $1<j<L$.
Such behavior is well known in both classical and quantum non-equilibrium open quantum system \cite{nrg_lindblad}.

The main parameter characterizing a stationary solution (after a relatively
short-time transient) is the average time $T$ that people spend from when they enter
the system till they exit. To evaluate $T$, we make a number of simplifying assumptions. Specifically, we
assume that  the (stationary) flow in either
direction does not depend on the direction itself and that the densities at any cell are independent of time
and uniform, that is, that $n_{\{ S , I , E \} , \lambda ; j}$ is independent of both $\lambda$ and $j$. As a result,
dropping the indices $j$ and $\lambda$ and making the other simplifying assumptions listed above, we see that
Eqs.~(\ref{list.1})-(\ref{list.3}) reduce to

\begin{eqnarray}
\frac{dn_{S}}{dt} & \approx & -\omega_{c}n_{S}n_{I}\left(1+\mu\right) \; , \nonumber \\
\frac{dn_{I}}{dt} & \approx & 0\; , \nonumber \\
\frac{dn_{E}}{dt} & \approx &  \omega_{c}n_{S}n_{I}\left(1+\mu\right)
\;\;\;\; ,
\label{ave.1}
\end{eqnarray}
\noindent
with $\mu=1$ for scenario A and B and $\mu=0$ for scenario C.

In solving Eqs.~(\ref{ave.1}), we assume $n_{\lambda;j}= n_{S,\lambda;j}+n_{I,\lambda;j}+n_{E,\lambda;j} = \bar{n}$
to be constant and independent of both $j$ and $\lambda$. Accordingly, $\bar{n}$ only depends on the rate of people
entering the system. The total number $N$  of individuals in the system at time $t$ is proportional to the
number of entering points, that is, to the number of different values of $\lambda$,  times the rate at which people enter
the system, $\omega_{\rm in}=\omega_{{\rm in};S,\lambda}+\omega_{{\rm in};I,\lambda}+\omega_{{\rm in};E,\lambda}$, minus the exit rate, $\omega_{\rm out}$ (which we  assume to be
equal to $\omega_h$,  times the total number of people in the last cells that, in the uniform, stationary
regime, is just equal to  $N/L$). Therefore, one obtains

\begin{equation}
\frac{dN}{dt}=   \omega_{\rm in} \Lambda-\omega_{h}\frac{N}{L}
\;\;\;\; ,
\label{totano}
\end{equation}
\noindent
with $\Lambda$ being the number of different values of $\lambda$. Eq.~(\ref{totano}) is
solved by setting

\begin{equation}
N (t) = \frac{\omega_{\rm in} \Lambda L }{\omega_{h}} \left( 1- e^{-\frac{\omega_{h}}{L}t}\right)
\:\:\:\: .
\label{purpetiello.1}
\end{equation}
\noindent
Extrapolating from Eq.~(\ref{purpetiello.1}) the asymptotic value of $N(t)$ for $t \to \infty$,
we can readily get the average  density of people in each cell for each value of $\lambda$, $\bar{n}$,
which is given by

\begin{equation}
\bar{n}=\frac{N}{ \Lambda L}=\frac{\omega_{\rm in}}{\omega_{h}}
\;\;\;\; .
\label{purpetiello.2}
\end{equation}
\noindent

Once $\bar{n}$ is fixed by the (asymptotic) system dynamics, it is
still possible for individuals in the  ``local'' population to switch among
compartments. This is, in fact, encoded in the ``local'' SEIR-like  Eqs.~(\ref{ave.1}).
To discuss the SEIR-dynamics, we therefore solve
Eqs.~(\ref{ave.1}) by setting $n_S ( t = 0 ) = n_{0 , S}$,  $n_I ( t = 0 ) = n_{0 , I}$,
and  $n_E ( t = 0 ) = n_{0 , E}$, with

\begin{eqnarray}
n_{0,S} &=& \delta_S \bar{n} \; , \nonumber \\
n_{0,I} &=& \delta_I \bar{n} \; , \nonumber \\
n_{0,E} &=& \left(1-\delta_S-\delta_I\right) \bar{n}
\:\:\:\: ,
\label{purptiello.3}
\end{eqnarray}
\noindent
and $0 \leq \delta_I , \delta_S \leq 1$, $\delta_I + \delta_S \leq 1$.
Solving Eqs.~(\ref{purptiello.3}), one eventually obtains

\begin{eqnarray}
n_{S}\left(t\right) & = & n_{0,S}e^{-\omega_{c}n_{0,I}\left(1+\mu\right)t} \; ,  \nonumber \\
n_{I}\left(t\right) & = & n_{0,I} \; , \nonumber \\
n_{E}\left(t\right) & = & n_{0,E}+n_{0,S}\left(1-e^{-\omega_{c}n_{0,I}\left(1+\mu\right)t}\right)
\:\:\:\: .
\label{purptiello.4}
\end{eqnarray}
\noindent
In our approach, the key  observable to quantify the level of infection due to
individual motion through our system is the infection ratio $C_{r , \lambda : j } ( t)$, that is,
the number of people that in cell $\left(j,\lambda\right)$ get infected in a time $t$ normalized to the
number of healthy people that entered the system at $t=0$. Clearly, within our stationary
solution we expect  $C_{r , \lambda : j } ( t)$ to be independent of both $\lambda$ and $j$.
Accordingly, for the sake of simplicity we henceforth denote it simply as $C_r ( t )$.
 From Eqs.~(\ref{purptiello.4}), we obtain that the infection ratio at time
$T$, $C_r$, is given by

\begin{equation}
C_{r}=\frac{n_{E}\left( T \right)-n_{0,E}}{n_{0,S}}=\left(1-e^{-\omega_{c}n_{0,I} \left(1+\mu\right) T }\right)
\;\;\;\; .
\label{purpetiello.5}
\end{equation}
\noindent
Apparently, $C_r$ measures the hazard for a healthy individual to go across the system in the
possible presence of infected people. Among the parameters at the right-hand side of
Eq.~(\ref{purpetiello.5}), $n_{0 , I}$ depends on the environmental conditions about
the infection spillover, $\mu$ depends in a known way on the system topology
(see the previous discussion) and, therefore, $T$ is the only parameter
that has to be estimated. While, in general, $T$ can be extracted from the results
of the numerical simulation, for $\omega_{\ell}=0$ it be analytically computed through a weighted average, as
we discuss next.

``Mimicking'', in a sense,   the cellular automata approach, to compute $T$,
we assume that the evolution in time of the system takes place via
a discrete sequence of elementary time steps, each one of duration $\Delta t$.
Accordingly, the fastest path taking an individual from the entrance to the
exit of the system has duration $T_{min} = L \Delta t$. In general, however, the topology
of the system   allows for backturns, which make the actual time
spent in the system larger than $T_{min}$. As a result, one obtains

\begin{equation}
T =\Delta t\sum_{M=0}^{\infty}\left(L+M\right)p_{h}^{L}\left(1-p_{h}\right)^{M}\left(\begin{array}{c}
L-1+M\\
M
\end{array}\right)
\;\;\;\; ,
\label{purpetiello.6}
\end{equation}
\noindent
with the probability $p_h = \omega_h \Delta t$.
The right-hand side of Eq.~(\ref{purpetiello.6}) corresponds to a weighted average,
with the weight given by the product of the probability to exit from the system in $\left(L+M\right)$ steps, that is $p_{h}^{L}\left(1-p_{h}\right)^{M}$,
times the number of permutation of such probability, that is $\left(\begin{array}{c} L-1+M\\ M \end{array}\right)$, where the  term $-1$ takes into account that the
probability string must finish with an hopping. From Eq.~(\ref{purpetiello.6}), one eventually finds

\begin{equation}
T  =\Delta t\frac{L}{p_{h}}=\frac{L}{\omega_{h}}
\;\;\;\; ,
\label{purpetiello.7}
\end{equation}
\noindent
that is independent of $\Delta t$, as expected.

\begin{figure}
\center
\includegraphics*[width=1 \linewidth]{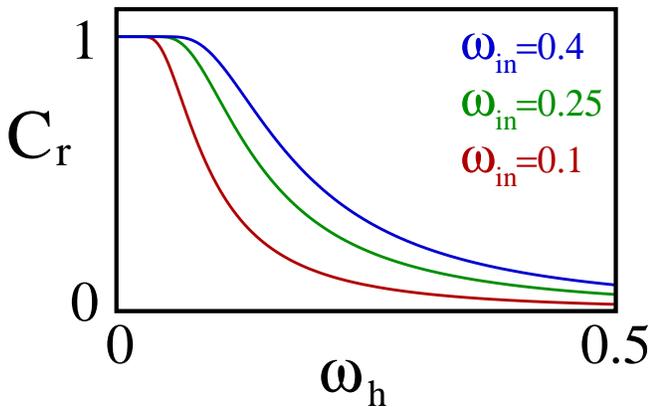}
\caption{$C_r$ as a function of $\omega_h$ (expressed in s$^{-1}$) for $\omega_{\ell}=0$ and, respectively, $\omega_{\rm in} = 0.1 {\rm \ s}^{-1}$ (red curve),
$\omega_{\rm in}=0.25 {\rm \ s}^{-1}$ (green curve), $\omega_{\rm in}=0.4 {\rm \ s}^{-1}$ (blue curve). The other parameters
are $L=50$, $\delta_E = 0$, $\delta_I = 0.05$, $\omega_c =0.025 {\rm \ s}^{-1}$, $\mu = 0$.}
\label{cr_ph}
\end{figure}

\begin{figure}
\center
\includegraphics*[width=1 \linewidth]{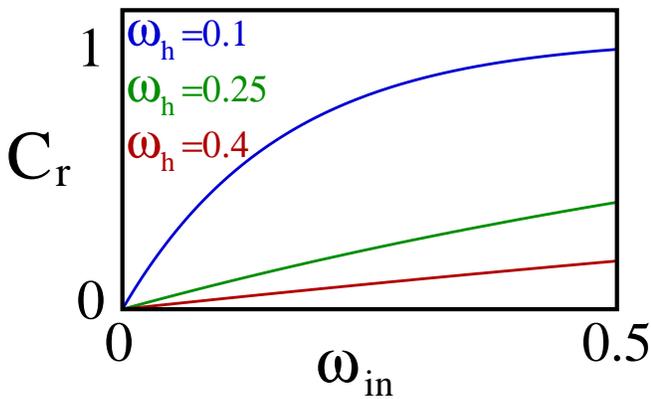}
\caption{$C_r$ as a function of $\omega_{\rm in}$ (expressed in s$^{-1}$) for $\omega_{\ell}=0$ and, respectively, $\omega_{h} = 0.4 {\rm \ s}^{-1}$ (red curve),
$\omega_{h}=0.25 {\rm \ s}^{-1}$ (green curve), $\omega_{\rm in}=0.1 {\rm \ s}^{-1}$ (blue curve). The other parameters
are $L=50$, $\delta_E = 0$, $\delta_I = 0.05$, $\omega_c =0.025 {\rm \ s}^{-1}$, $\mu = 0$.}
\label{cr_pin}
\end{figure}

As long as $\omega_\ell = 0$, inserting  Eq.~(\ref{purpetiello.7}) into Eq.~(\ref{purpetiello.5}) for
$C_r$, one gets
\begin{equation}
\label{cierre}
C_r=1-\exp \left [- \delta _I \frac{\omega _c \omega _{in}}{\omega _h^2} (1+\mu) L \right ]
\;\;\;\; ,
\end{equation}
and then one can draw plots of the infection ratio as a function of either $\omega_h$, at a given $\omega_{\rm in}$, or of
$\omega_{\rm in}$, at a given $\omega_h$. In Fig.~\ref{cr_ph}, we draw  $C_r$ as a function of $\omega_h$ for three sample
values of $\omega_{\rm in}$. Remarkably, we see that, at large enough hopping rate between neighboring lattice cells,
$C_r$ barely depends on $\omega_{\rm in}$ and keeps as low as $1\% - 10\%$. At variance, in Fig.~\ref{cr_pin} we
 we draw  $C_r$ as a function of $\omega_{\rm in}$ for three sample   values of $\omega_h$. Here, we see that
the dependence of $C_r$ on $\omega_{\rm in}$ is strongly affected by the value of $\omega_h$. In particular, while
keeping $\omega_{h}$ as large as 0.4 s$^{-1}$ allows for maintaining $C_r$ of the order of 10\%, even
at pretty large values of $\omega_{\rm in}$, at variance, as soon as $\omega_h$ becomes of the order of
0.25 s$^{-1}$, $C_r$ rises up to 40\% when $\omega_{\rm in} \sim 0.5 {\rm \ s}^{-1}$. Finally, for
$\omega_h =  0.1{\rm \ s}^{-1}$, we see that $C_r$ is  $\sim$ 50\% already  for $\omega_{\rm in}
\sim  0.15{\rm \ s}^{-1}$. Thus, from the synoptic comparison of the plots in  Fig.~\ref{cr_ph} and Fig.~\ref{cr_pin},
one may on one hand  infer how  $C_r$ is poorly sensitive to the
value of $\omega_{\rm in}$, provided $\omega_h$ is large enough, which basically
implies that, if the individual flow through the aisles is made fast enough, the
entry rate of people into the system from the outside is not a crucial parameter
to keep $C_r$ low. On the other hand, one finds that, as a function of $\omega_{\rm in}$,
$C_r$ is strongly sensitive to the value of $\omega_h$. Indeed,  Fig.~\ref{cr_pin}
basically shows how  easy is for $C_r$ to become already as large as 50\%, if  $\omega_h$ is not kept large enough to avoid crowding within the system.

\noindent

\begin{figure}
\center
\includegraphics*[width=1 \linewidth]{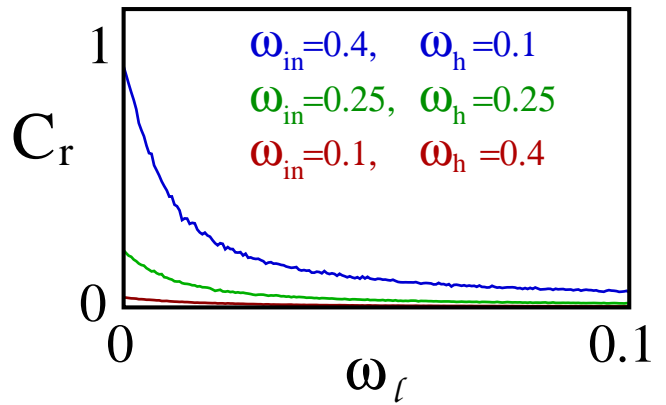}
\caption{$C_r$ as a function of $\omega_{\ell}$ (expressed in s$^{-1}$) and, respectively, $\omega_{\rm in} = 0.1 {\rm \ s}^{-1},
\omega_{h} = 0.4 {\rm \ s}^{-1}$ (red curve),
$\omega_{\rm in} = 0.25 {\rm \ s}^{-1},
\omega_{h} = 0.25 {\rm \ s}^{-1}$  (green curve), $\omega_{\rm in} = 0.4 {\rm \ s}^{-1},
\omega_{h} = 0.1 {\rm \ s}^{-1}$  (blue curve). The other parameters
are $L=50$, $\delta_E = 0$, $\delta_I = 0.05$, $\omega_c =0.025 {\rm \ s}^{-1}$,  $\mu = 0$.   }
\label{tmedio}
\end{figure}

\noindent
We discuss now the effects on $C_r$ of allowing individuals to change their direction of motion, once in
the system. On intuitive grounds, one expects that letting $\omega_\ell \neq 0$ should lower $C_r$, at fixed values of
the other system parameters.  This is reasonable because   now individuals are allowed  to exit the system from the same
side they entered. This means that an individual who enters the system from say cell $\left(1,1\right)$ and who wants to reach, for example, cell $\left(1,2\right)$ (on the other side of the street), does no
more need to run across all the street till the cell with $j=L$ and back, thus consistently lowering the risk of infecting other people,
or of being  infected by other people meanwhile. In fact, allowing people to change their walking direction allows them to reach faster the
shop they are interested in by lowering the path accordingly. On the mathematical side, having $\omega_\ell \neq 0$ does no
more allow for exactly computing $T$ by means of a procedure similar to the one leading to Eq.~(\ref{purpetiello.6}).
Therefore, to plot $C_r$ as a function of $\omega_\ell$ with the other system parameters being fixed,
we numerically derived $T$ at given $\omega_\ell$ and therefore substituted the corresponding value in Eq.~(\ref{purpetiello.5})
for $C_r$. In Fig.~\ref{tmedio}, we report the corresponding curves for $C_r$ as  functions of $\omega_\ell$, with the other
system parameters set as discussed in the caption. Apparently, we see that increasing $\omega_\ell$ always acts to
lower $C_r$. Also, as we already inferred from the plots in  Fig.~\ref{cr_ph} and Fig.~\ref{cr_pin}, we see that $C_r$ is
further lowered by simultaneously increasing $\omega_h$ and lowering $\omega_{\rm in}$, as we extensively  discussed above.

\begin{figure}
\center
\includegraphics*[width=1 \linewidth]{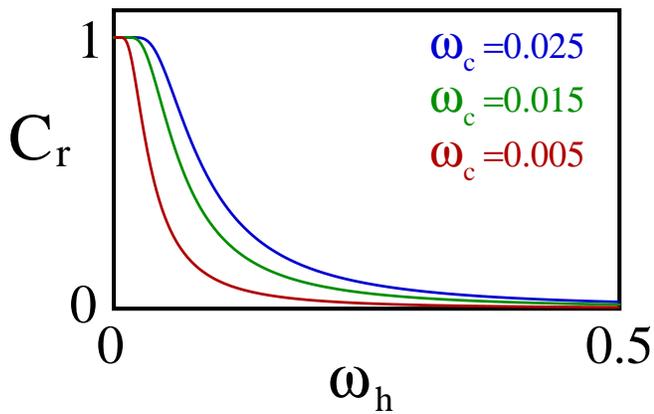}
\caption{$C_r$ as a function of $\omega_{h}$ (expressed in s$^{-1}$) and, respectively, $\omega_{\rm in} = 0.1 {\rm \ s}^{-1}$,
$\omega_\ell = 0$, $\omega_{c} = 0.005 {\rm \ s}^{-1}$ (red curve),
$\omega_{c} = 0.015 {\rm \ s}^{-1}$  (green curve), $\omega_{c} = 0.025 {\rm \ s}^{-1}$  (blue curve). The other parameters
are $L=50$, $\delta_E = 0$, $\delta_I = 0.05$,  $\mu = 0$.}
\label{conta}
\end{figure}

\noindent
Finally, to investigate how, and to what extent, $C_r$ depends on a tunable parameter, in Fig.~\ref{conta} we plot
$C_r$ as a function of $\omega_h$ for different values of the infection rate $\omega_c$. Indeed, $\omega_c$ is a
parameter one can effectively act on from the outside by, for instance, letting people entering the system to wear
a mask and/or to keep social distancing, {\it et cetera}. From the plots of Fig.~\ref{conta}, as expected, we see that while 
at a given $\omega_h$, lower $\omega_c$ and lower $C_r$ correspond, at the
lowest value of $\omega_c$, one sees that $C_r$ stays lower than 10\% as soon as $\omega_h \geq 0.1 {\rm \ s}^{-1}$.
Apparently, this is a quantitative evidence of the effectiveness of using as many measures to prevent infection as possible,
such as masks and social distancing. A good combination of prevention measures with a pertinent engineering of the
individual pathways inside a given system, as well as with an appropriate regulation of the entrance rate in the system,
can apparently work as an effective mean to keep the level of infection pretty low.

\noindent
To comment about our result, we note that our  theoretical model is definitely able to catch some interesting qualitative behaviors.
However, it overestimates the infection  ratio, due the mean field approximation we employ to provide an explicit form for
 the joint probability function, $f_{\lambda;j}$. Indeed, for example, if we consider scenario C with a hopping probability equal to  one, the contagion
should be exactly zero if less than an individual  enters the system at each turn. However, the theoretical model is not able to reproduce this result
because it totally neglects the actual people distribution in real space within the system. Furthermore, the mean field approximation is not able to properly
distinguish between scenarios B and C. Indeed, as stated above, implementing the mean field approximation fixes the
parameter $\mu$ at $\mu=1$ in scenario B and at $\mu =0$ in scenario C. Yet, from   the formula for $C_r$ within mean field approximation,
Eq.~(\ref{purpetiello.5}), we see that the plots in the two cases just collapse onto each other, provided one sets
 $ \omega_{\rm in}$ in scenario C to be twice as large as $\omega_{\rm in}$ in scenario B. This is a consequence of the fact that, in both cases, people
are divided into two separate groups: in scenario B they are divided into two lanes; in scenario C they are divided according to their
direction of motion.  In both cases they interact with half of the people they would interact with in scenario A. However, in scenario C,
we have to consider the relative speed between people that, in the mean field approximation is simply thrown away (two people with
opposite velocity  meet for sure while people moving in the same direction do not).
For this reason, as well as  defining a playground to extend our approach to  a systematic analysis of the  transient regimes and of a generic case of time-space dependent
rates, in the following  we resort to the cellular automata approach, by means of which we will be able to
implement the joint probability function in terms of simple rules.
It is worth noting that, in general, the CA rules can be traded for an effective set of Fokker-Planck equations to directly model the time 
evolution of the on-site probability density \cite{fokker}. However, the resulting model can hardly be solved by analytical method and is more computational demanding than the CA approach.

\section{The cellular automata rules and their implementation}
\label{rule}

In this section we discuss in detail the rules of the cellular automata  describing the spread of an infection within an open,
finite connected graph and how we implement them. In particular, to keep consistent with the
analytical derivation of Sec.~\ref{theo}, in the following  we focus on the three different scenarios we proposed there.

Let us begin our discussion with  scenario A. Referring to Fig.~\ref{CA}, we model this system as a $2\times L$ square grid
with von Neumann neighborhood. On indexing each plaquette of the lattice with a ``row'' and a ``column`` index, we
use the row index to store the information on  the direction of the motion of each individual, while  the column index keeps
track of the spatial position. Accordingly, we regard each cell as a portion of a ``road'' of physical length $L$.
The first row of the matrix represents people moving to
the left (that corresponds to having $\lambda=-1$ in the notation of Sec.~\ref{theo}), from cell $(1,L)$ to cell $(1,1)$. At
variance, the  second row represents people  moving to the right (corresponding to $\lambda=+1$  in the notation of Sec.~\ref{theo}),
from cell $(2,1)$ to cell $(2,L)$ (note that, in scenario A, cells $(1,j)$
and $(2,j)$ overlap each other in real space). To each cell $(\lambda , j)$, we associate three positive integers, $n_{S,\lambda;j}$, $n_{E,\lambda;j}$
and $n_{I,\lambda;j}$, respectively corresponding to the number of healthy, exposed and infectious individuals moving in the  direction $\lambda$, as
defined in Sec.~\ref{theo}.
While we allow more than a single individual to occupy the same cell, to take into account that each cell corresponds to a finite
region in space, we put a constraint on  maximum limit of individuals in each
cell. Letting $d$ to be the side of each (square) cell, we are therefore limiting the maximum number of people that can physically enter a $d \times d$
square region of space.  Accordingly, we require that, $\forall j = 1 , \ldots , L$, we obtain $\sum_{\lambda=\pm 1} (n_{S,\lambda;j}+n_{E,\lambda;j}+n_{I,\lambda;j})\leq n_{\rm{max},j}$,
with $n_{\rm{max},j} = n_{\rm{max}} = 15$ for all cells, as it appears to be reasonable for the cell size we consider (see below for
the detailed discussion about our choice of the system parameters).

The populations inside each cell are updated every time step $\Delta t$.
Each time step is composed by two phases: the {\it movement turn} and the {\it contagion turn}.
At variance with respect to our theoretical framework in Sec.~\ref{theo}, in this
case we are considering {\it integrated rates} at each single turn, that is, probabilities.
Therefore, referring to the definition of the various rates in Sec.~\ref{theo}, we
denote with $p_h = \omega_h \Delta t$ the  probability that, in a single turn,  an individual either moves backward
from cell $(1,j)$ to  cell $(1,j-1)$, or it moves forward from cell $(2,i)$ to cell $(2,j+1)$ (focusing on the
``inner'' cells, that is, $1<j<L$). Going along the rate formalism of Sec.~\ref{theo},  we
define $p_\ell = \omega_\ell \Delta t$ to be the probability for an individual to change in a turn its direction of
motion, that is, to  move from cell $(1,j)$ to cell $(2,j)$, or {\it vice versa}.
Individuals, whether healthy, exposed or contagious,  enter the system at each turn from the boundary cells   $(1,L)$ and $(2,1)$.
In particular, letting  $p_{\rm in} = \omega_{\rm in} \Delta t$  the probability that one individual enters the system in a time step $\Delta t$ and
also letting  $\delta_S$, $\delta_E$, $\delta_I$ the average fraction of the total population corresponding to
healthy, exposed and infectious individuals respectively, we have that the probability for  a
healthy,  exposed, or infectious individual to enter the system in a single time step is respectively given by
 $p_{{\rm in},S}=\delta_S p_{\rm in}$, $p_{{\rm in},I}=\delta_I p_{\rm in}$ and $p_{{\rm in},E}=\delta_E p_{\rm in}$, clearly   with $\delta_S+\delta_E+\delta_I =1$
(note that, while  $p_{\rm in}$ is the entrance probability for a single pedestrian, in general, more than a pedestrian can enter the system
at each turn). Finally, individuals can exit the system from the cells $(1,1)$ or $(2,L)$.
When a pedestrian  exits the system, it is removed from the total count of people within the lattice and, at
the same time, it  triggers a counter that keeps trace of its health status. Likewise, another counter keeps trace of the people
that effectively entered the system. Each hopping, entering or change of direction is allowed only if the total population inside the target cell
has not saturated to the allowed value $n_{\rm{max}}$.

The infection turn takes place right after the movement turn. In each cell $(\lambda,j)$, the health status of each pedestrian has a
probability $p_c = \omega_c \Delta t$ to switch  from $S$ to $E$ for each contagious individual that is present in cells $(1,j)$ and $(2,j)$.
Consistently with the assumptions discussed in Sec.~\ref{theo} and in Appendix \ref{rseir}, we do not allow the status of $E$- and $I$-individuals
to change.

Switching to scenario B, one readily sees that it  is a simple ``double copy'' of scenario A. In this case, the matrix has four rows,
rather than two, with two rows per each ``side of the street''. Therefore, one has two different directions of motion per each
side of the street,   with the possibility, for a single individual, to switch, at the same time walking side and/or direction.
In order to compare the results between scenarios A and B we keep fixed the incoming individual flow halving the incoming probability $p_{\rm in}$.

Scenario C is the same as scenario A, except for the fact that people moving
 in opposite directions  are physically separated in real space.
Therefore, during the  contagion turn, an $S$-individual    can only be infected  by an $E$-individual moving in the same direction. Furthermore,
the constraint on the maximum allowed number of individual in each cell must be satisfied separately for each value of $\lambda$, that is,
the constraint takes the form  $(n_{S,\lambda;j}+n_{E,\lambda;j}+n_{I,\lambda;j})\leq n_{\rm{max},\lambda}$.

At the end of the simulation, we compute the infectious ratio $C_r$ as the number of $E$-individuals leaving
the system, minus the number of individuals already exposed before entering into the system,
over the number of healthy individuals entered into the system. 
In particular, to compute $C_r$  we stop  the simulation at a time t=7h, that correspond approximately to a shop 
working day. As long as the simulation time  is greater than the average time   that people spend from when they enter 
the system till they exit, the system has already reached a non equilibrium stationary state and the effects associated 
to the stochastic nature  of the system are mediated to zero, as a consequence of the ergodic hypothesis.

\begin{figure}
\center
\includegraphics*[width=1. \linewidth]{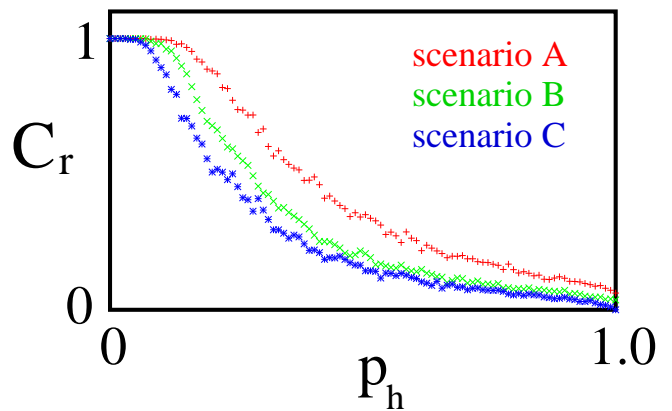}
\caption{$C_r$ as a function of $p_h$ with, respectively, $p_{\rm in}=0.5$, $p_{\ell}=0$, $p_c=0.05$, $\delta_E=0$,
$\delta_I=0.05$ computed within CA approach for scenario A (red pointplot), scenario B (green pointplot), scenario
C (blue pointplot). As expected, scenario A is the worst, since individuals
moving toward opposite directions are not spatially separated from each other, scenario C is
the best, scenario B is halfway between the two of them \cite{note}.  }
\label{CA_cr_ph}
\end{figure}

The cellular automata parameters depend on the model and on the physical system we are interested to describe.
In our simulation, we employ the following parameters:

\begin{itemize}
  \item {\it The cell dimension $d$:} this should be pertinently  chosen to be of the order of the safety distance between individuals, to keep consistent with
  our assumption that infection spread only happens inside a cell. Accordingly, we set $d=2$m;

  \item {\it The number of cells $L$:} this is not a crucial parameters. It measures the length of the pathway we are considering in units of
  $d$. Throughout all out calculations, we set $L=50$ but, as stated above, it can be readily changed to simulate longer, or shorter,
  paths;

  \item {\it The time step $\Delta t$:} this measure the time required by a pedestrian to walk for $d$ meters without slowing down.
 In the following, we set $\Delta t=2$s;

 \item {\it The infection probability $p_c$:} in general, this depends  on the effective virus transmission probability, on
the pedestrian health status (which, in turns, depends on age, gender, {\it et cetera}),
and by the  protective equipment wore by individuals, such as  masks and  gloves;

\item {\it Percentage of infectious individuals effectively entering the system, $\delta_I$:}
while, in principle, $\delta_I$ is determined by  the average percentage of infectious people
to the main population, at the entrance to the system it can be strongly  reduced (compared to the outside) by, {\it e.g.},
checking the body temperature at the entrance and by forbidding people with symptoms like fever or cough to enter the system.
 Indeed, in the case of infection by Covid-19, it has been estimated that about $43,8\%$ of infectious people have fever before  hospitalization. Other discriminants, for checks at
the entrance,  can be age and gender, indeed the median age was computed as 47 years, $58.1\%$ were males, and only $0.9\%$
of patients were aged below 15 years~\cite{Covid5}.

  \item {\it Entrance probability $p_{\rm in}$, ``hopping'' probability without changing direction,
  $p_h$, and probability of changing direction, $p_\ell$}: these   depend  on the number of individuals in the system
  and on the time spent by them in units of $\Delta t$. Realistic estimates for those probabilities (or
  for the corresponding rates)  can be extracted from the crowd fluxes measured,
  for instance,  by the transit crowdedness function of {\it Google Maps}, in previous years. In addition, one should
 also take into account the possibility of ``artificially modifying'' these probabilities by, {\it e.g.},
influencing people with  disclaimers, turnstiles, watchmen or "nudges"~\cite{nudge}.
\end{itemize}
As a main sample of CA results, in Fig.~\ref{CA_cr_ph} we plot $C_r$ as a function of
$p_h$ for all the  three scenarios described in Sec.~\ref{theo}. To make the windows of values of
the independent variable consistent with the one we use to draw Figs.~\ref{cr_ph},\ref{conta}, we let
$0 \leq p_h \leq 1$ which, given the relation $p_h = \omega_h \Delta t$ and since
$\Delta t =2$s, corresponds to $0 \leq \omega_h \leq 0.5 {\rm \ s}^{-1}$ of Figs.~\ref{cr_ph},\ref{conta}.
As expected from the main features of the three scenarios, at values of all the system parameters all equal
in the three different cases, for what concerns the infection rate, scenario A is the worst, since individuals
moving toward opposite directions are not spatially separated from each other, scenario C is
the best, due to the condition $\mu_{\lambda , \bar{\lambda}} = 0$ (see the discussion in
Sec.~\ref{theo} for details), scenario B is halfway between the two of them. Remarkably,  while we recover an
acceptable qualitative agreement with the results discussed in Sec.~\ref{theo},
we observe how, differently from the mean field approximation,  the CA approach is able to correctly discriminate between scenarios
B and C and to reproduce the right limit $C_r \to  0$ limit for $p_h \to 1$.

\begin{figure}
\center
\includegraphics*[width=1. \linewidth]{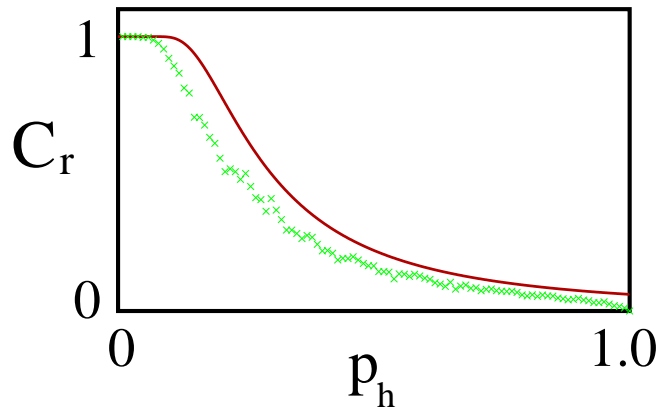}
\caption{$C_r$ as a function of $p_h$ computed in scenario C (green pointplot) and within
mean field approximation with $\mu = 0$ (full red line) with  $p_{\rm in}=0.5$, $p_{\ell}=0$, $p_c=0.05$, $\delta_E=0$,
$\delta_I=0.05$. Apparently, the mean field calculation always overestimates $C_r$, compared to the ``exact'' CA
result.}
\label{compar}
\end{figure}

To ease the comparison  between the CA results and the (approximate) analytical one, to check the reliability of the approximations
we employed along the derivation of Sec.~\ref{theo}, in drawing the plots in Fig.~\ref{CA_cr_ph} we chose the parameters
so that, once all the probabilities are converted into rates by dividing all of them by $\Delta t$, they exactly
correspond to the ones we used to draw Fig.~\ref{cr_ph}. In particular,  Fig.~\ref{cr_ph} was drawn for $\mu = 0$, which
correctly describes scenario C only. For this reason, in Fig.~\ref{compar} we draw a synoptic plot of the
curve of  Fig.~\ref{cr_ph}  corresponding to the parameters chosen to draw Fig.~\ref{CA_cr_ph} and of the points
of Fig.~\ref{CA_cr_ph} corresponding to scenario C (note that we use $p_{\rm in}$ as the sole independent variable for
both plots, after converting $\omega_{\rm in}$ into $p_{\rm in}$ using $p_{\rm in} = \omega_{\rm in} \Delta t$). Importantly enough,
the synoptic comparison shows that, at given system parameters, the mean field approximation systematically overestimates
$C_r$ at a given $p_h$, which shows that the simple analytical approach, in a sense, provides in a simple way a
``safe'' upper bound on the infection risk.

\noindent

\begin{figure}
\center
\includegraphics*[width=1. \linewidth]{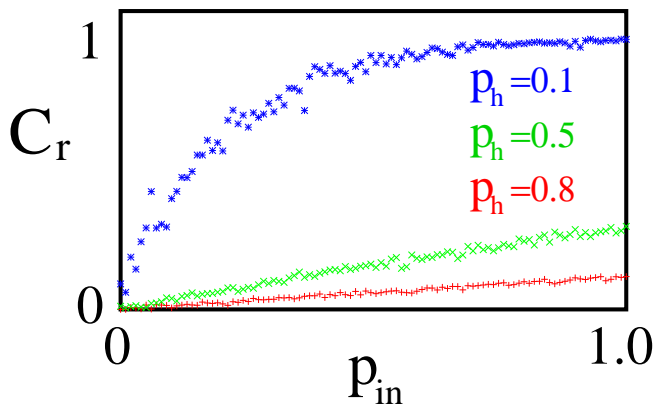}
\caption{$C_r$ as a function of $p_{\rm in}$ and, respectively, $p_{\ell}=0$, $p_c=0.05$, $\delta_E=0$, $\delta_I=0.05$ and $p_{h}=0.8$ (red pointplot),
$p_{h}=0.5$ (green pointplot), $p_{h}=0.1$ (blue pointplot), for scenario C.}
\label{CA_cr_pin}
\end{figure}

\noindent
To further compare the CA approach to the analytical mean field approximation, in Fig.~\ref{CA_cr_pin} we plot
the numerical data from the CA for $C_r$ as a function of $p_{\rm in}$ at $p_\ell = 0$ for various values of
$p_h$ and with all the other parameters chosen as in the derivation of Sec.~\ref{theo} (see the figure caption
for details). Qualitatively speaking, we see that the trend of the data in  Fig.~\ref{CA_cr_pin} is the
same as we display in Fig.~\ref{cr_pin}, which was derived by applying the mean field approximation to
the generalized SEIR model. Specifically, at a given value of $p_{\rm in}$, increasing the probability
for individuals to move from one cell to the neighboring one (that is, increasing the average speed
of the pedestrian motion in the pathway) determines a remarkable lowering of $C_r$, as expected
in view of the fact that, as discussed in the previous section, due the possibility for
individuals to exit from the same side of the street, the mean time spend by each of them inside
the system is reduced, as well as the probability of  infecting,  or of being infected.

\begin{figure}
\center
\includegraphics*[width=1. \linewidth]{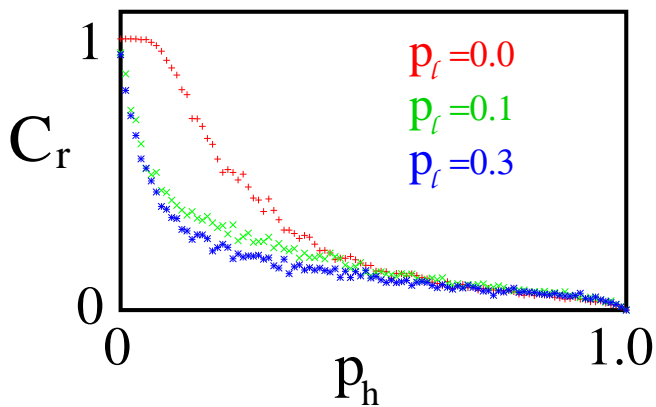}
\caption{$C_r$ as a function of $p_h$ and, respectively, $p_{\rm in}=0.5$, $p_c=0.05$, $\delta_E=0$, $\delta_I=0.05$ and $p_{\ell}=0$ (red pointplot),
$p_{\ell}=0.1$ (green pointplot),  $p_{\ell}=0.3$  (blue pointplot), for scenario C.}
\label{CA_cr_ph_pl}
\end{figure}

Finally, to complement the results reported in Fig.~\ref{conta} with the corresponding analogs
derived within CA framework, in Fig.~\ref{CA_cr_ph_pl} we plot $C_r$ as a function of $p_h$
for scenario C, for various values of $p_\ell$ and all the other parameters set to quantitatively
ground the comparison with Fig.~\ref{conta} (see the figure caption for details).
As expected, the CA results confirm that increasing $p_\ell$ by keeping all the other parameters
fixed acts to lower $C_r$, as it basically lowers the average time spent by individuals in
the system (see Sec.~\ref{theo} for a detailed discussion about this point).

\noindent

\noindent
Eventually, we recover an excellent consistency between the (approximate) analytical results and
the (basically exact) numerical ones obtained within the CA framework. This evidences that, on
one hand, the mean field approach we employ to solve the generalized SEIR equation provides
an acceptable level of qualitative description of the system behavior, on the other hand,
that the CA approach can be effectively implemented to improve the quantitative
reliability of the results, when necessary.
It is worth stressing that, in addition to the mean field effect, another 
cause of discrepancy between the deterministic SEIR model and the stochastic CA simulation 
can emerge  due the fact that, while in the numerical simulation pedestrians are subject to both mobility and contagion at each turn, 
in the differential equations there is no term that simultaneously involves mobility and contagion. Eventually, 
for a better comparison between analytical and numerical results, an adapted version for time-varying networks 
of the Gillespie algorithm for stochastic methods can be implemented \cite{gillespie0, gillespie1}.

\section{Conclusions}
\label{conclusion}

In this  paper we have formulated a generalized SEIR model on a graph that allowed  us to
describe the population dynamics of an open crowded place with an, in principle,
arbitrary topology. To illustrate the effectiveness of our model, we have
discussed a few, simple paradigmatic cases, which we have treated both analytically, within a mean field approach to
the full set of SEIR model differential  equations, and numerically, by means of a cellular automata   simulation of the
individual dynamics in the system. As a main result of our derivation,
we were able to provide the infection ratio  $C_r$  as  a function of ``tunable'' system parameters,
which eventually enables us to show to what extent controlling
human-depending effects may act to lower the disease spread in the system.

As an immediate further development of our work, we note that, within our approach, one may readily
extrapolate the ratio between individuals that become exposed inside the system and infectious
individual coming from outside as

\begin{equation}
R = \frac{\delta_S}{\delta_I}C_r
\;\;\;\; .
\label{R-def}
\end{equation}
\noindent
Dividing $R$, obtained within CA simulation, by the time by which we run the simulation and then multiplying the result   by the overall
fraction of time usually spent by  an individual into the system under analysis, in a period as long as the incubation time,
one may obtain the number of people infected by each infectious
individual, that is the basic reproduction number $R_0$~\cite{R0}, related to a specific environment. As a next development of our
work, we plan to estimate this quantity, that is only a fraction of the cumulative $R_0$ (that is the sum of the $R_0$ of all the places the individual spent time in),
for different scenarios, {\it e.g.},  a shopping center, a pedestrian track, a gym, a school, and so on. Eventually, we plan to use it our results as a tool to evaluate the infection
hazard of a given place compared to others and to, {\it e.g.},  suggest which place would be safer to open first after a global lockdown.
Finally, it is worth stressing that, as  $R_0$ is an effective parameter that depends on the virus and on the
social behavior, our approach allows for discriminating  the contribution due the virus intrinsic properties~\cite{Covid1,Covid2,Covid3,Covid4}
from  the human-depending effects like social distancing, personal protective equipment and so on, and to provide
a quantitative information about how to act to reduce the latter contribution.

Finally, we   remark that, while, in order to make the presentation of our approach the  straightest possible, we
confined ourselves to three simple sample scenarios, our approach can be easily generalized to more realistic and,
unavoidably, more complex graphs, by means of a proper implementation of the neighborhood between cells within CA simulation.
The rules that describe the human mobility, and in general people behavior, are strongly
scenario depending. In our model we did not consider a direct interaction between pedestrian, with individuals 
walking independently. In the high-density regime it would be interesting to introduce, for example, a more 
realistic description between pedestrian flows in different direction, as depicted in Ref.[\onlinecite{social_force}].  Indeed, as a
further development of our work,  we are planning to apply our approach to some specific and realistic cases for
which there would be available experimental data (entering/exit rate, mean time spent within the system, etc.),
so to eventually be able to perform a comparison between real results and estimates of our model. A particularly interesting 
aspect is to consider the addition of “attractive” or “trap” sites (popular shops, cafés, escalators), corresponding to local 
decreases of the hopping rate, that could represent high-risk transmission sites, in order to estimate how much they affect $R_0$.

\vspace{0.5cm}

{\bf Acknowledgements --}
A. N. was financially supported  by POR Calabria FESR-FSE 2014/2020 - Linea B) Azione 10.5.12,  grant no.~A.5.1.
D. G. and M. R. acknowledge  financial support  from Italy's MIUR  PRIN projects TOP-SPIN (Grant No. PRIN 20177SL7HC).
 
\appendix

\section{Review of the compartmental SEIR model}
\label{rseir}

The compartmental SEIR model is a generalization of the widely used SIR model in epidemiology~\cite{kam_1,kam_2}. Specifically,  the
SEIR model is based on parting a population of $N$ individuals (populating an isolated area, so that $N$ is assumed to be constant) into
four compartments, that is

\begin{itemize}
\item The {\it Susceptible}  compartment ($S$), that is made out of healthy individuals who can be affected by the contagion;
\item The {\it Exposed}  compartment ($E$), that is made out of individuals who have been infected, but are not yet infectious;
\item The {\it Infectious}  compartment ($I$), that is made out of infectious individuals (the ones that can infect individuals in the
$S$ compartment);
\item The {\it Recovered}  compartment ($R$), that is made out of individuals who either recovered from the infection, or  died.
\end{itemize}

In its standard and simplest formulation, the SEIR model describes the evolution in time of the number of individuals in each sector
by means of a set of differential, rate equations, given by

\begin{eqnarray}
\frac{ d S}{d t } &=& - \frac{ \omega}{N} S I \nonumber \\
\frac{ dE}{d t} &=&  \frac{ \omega}{N} S I - \alpha E \nonumber \\
\frac{ d I}{d t } &=& \alpha E - \gamma I \nonumber \\
\frac{ d R}{d t } &=& \gamma I
\:\:\:\: .
\label{appe.1}
\end{eqnarray}
\noindent
As discussed in the main text, we describe each lattice cell as a single SEIR model in which, however,
the total number of individuals can change in time, due to the nonzero probability of hopping between a cell
and the nearest neighboring ones. In particular, this implies that, even when considering stationary solutions
to the dynamical evolution equations for the local population, they are only on the average (in time) consistent with
the conservation of the total number of individuals per each cell. The various rates in Eqs.~(\ref{appe.1}) are clearly
identified as follows:

\begin{itemize}
\item $\omega$ is the ``specific infection rate'', that is, the probability per unit time that an individual belonging to $S$ is
infected by getting close to another individual belonging to $I$;
\item  $\alpha$  is the probability per unit time that an individual switches from $E$ to $I$, that is, the rate that the
virus incubation ends and the individual becomes infectious;
\item $\gamma$ is the rate for an individual from $I$ to switch to $R$. According to the specificities of the model,
$\gamma$ is identified with the healing-plus-death rate for an individual from $E$.
\end{itemize}

Even a rather simplified set of equations such as the ones in Eqs.~(\ref{appe.1}) can be able to provide reliable
informations on the infection spillover, provided the various rates at the right-hand side of Eqs.~(\ref{appe.1})
(the ``parameters'') are pertinently estimated. For instance, in the case of Covid-19 infection
spreadout in  Italy, we employ the numerical values for the parameters rigorously estimated in  Ref.~[\onlinecite{rig}], that
is,  $\omega = 2.25$[day]$^{-1}$, $\alpha=0.33$[day]$^{-1}$,    $\gamma=0.50$[day]$^{-1}$, values that appear to fit well the infection dynamics in
two among the mostly populated regions in Italy: Lombardy (Northern Italy) and Campania (Southern Italy).

While, over a long enough time, the system of Eqs.~(\ref{appe.1}) always implies, given the parameters
listed above, a maximum in the contagion spreading curve and an asymptotic saturation of $ R (t)$ to
$R_\infty = R ( t \to \infty ) = N$, throughout our work we set $\alpha = \gamma =0$ and, accordingly,
we neglect the equation for $R ( t )$. Indeed, in the specific system we describe, we assume that the time spent
by each individual inside the system is pretty short compared to the virus incubation and healing-plus-death rate.
Accordingly, $I ( t )$ can become different from  zero only because of people that come from outside, with a negligible
change due to the switch  in time  from  $E ( t )$ to $I ( t )$. Given the estimate of the parameter $\alpha$ we provide above,
the variation in $I ( t )$ over time intervals as long as a few hours is typically negligible and we
accordingly neglect it, by setting $\alpha$ to  zero from the very beginning and approximating
$ I ( t ) \approx I ( 0 )$. This eventually motivates assuming $R ( t ) = 0$, as well, as we
do throughout our derivation. Apparently, the above assumptions eventually lead to Eqs.~(\ref{ave.1}),
in the specific case corresponding to taking $\mu = 0$ in Eq.~(\ref{truly.2}), as we discuss in
detail in the main text of the paper.

\bibliography{pedestrian_biblio}

\begin{thebibliography}{49}%
\makeatletter
\providecommand \@ifxundefined [1]{%
 \@ifx{#1\undefined}
}%
\providecommand \@ifnum [1]{%
 \ifnum #1\expandafter \@firstoftwo
 \else \expandafter \@secondoftwo
 \fi
}%
\providecommand \@ifx [1]{%
 \ifx #1\expandafter \@firstoftwo
 \else \expandafter \@secondoftwo
 \fi
}%
\providecommand \natexlab [1]{#1}%
\providecommand \enquote  [1]{``#1''}%
\providecommand \bibnamefont  [1]{#1}%
\providecommand \bibfnamefont [1]{#1}%
\providecommand \citenamefont [1]{#1}%
\providecommand \href@noop [0]{\@secondoftwo}%
\providecommand \href [0]{\begingroup \@sanitize@url \@href}%
\providecommand \@href[1]{\@@startlink{#1}\@@href}%
\providecommand \@@href[1]{\endgroup#1\@@endlink}%
\providecommand \@sanitize@url [0]{\catcode `\\12\catcode `\$12\catcode
  `\&12\catcode `\#12\catcode `\^12\catcode `\_12\catcode `\%12\relax}%
\providecommand \@@startlink[1]{}%
\providecommand \@@endlink[0]{}%
\providecommand \url  [0]{\begingroup\@sanitize@url \@url }%
\providecommand \@url [1]{\endgroup\@href {#1}{\urlprefix }}%
\providecommand \urlprefix  [0]{URL }%
\providecommand \Eprint [0]{\href }%
\providecommand \doibase [0]{http://dx.doi.org/}%
\providecommand \selectlanguage [0]{\@gobble}%
\providecommand \bibinfo  [0]{\@secondoftwo}%
\providecommand \bibfield  [0]{\@secondoftwo}%
\providecommand \translation [1]{[#1]}%
\providecommand \BibitemOpen [0]{}%
\providecommand \bibitemStop [0]{}%
\providecommand \bibitemNoStop [0]{.\EOS\space}%
\providecommand \EOS [0]{\spacefactor3000\relax}%
\providecommand \BibitemShut  [1]{\csname bibitem#1\endcsname}%
\let\auto@bib@innerbib\@empty
\bibitem [{\citenamefont {Kermack}\ \emph {et~al.}(1927)\citenamefont
  {Kermack}, \citenamefont {McKendrick},\ and\ \citenamefont {Walker}}]{SIR0}%
  \BibitemOpen
  \bibfield  {author} {\bibinfo {author} {\bibfnamefont {W.~O.}\ \bibnamefont
  {Kermack}}, \bibinfo {author} {\bibfnamefont {A.~G.}\ \bibnamefont
  {McKendrick}}, \ and\ \bibinfo {author} {\bibfnamefont {G.~T.}\ \bibnamefont
  {Walker}},\ }\bibfield  {title} {\enquote {\bibinfo {title} {A contribution
  to the mathematical theory of epidemics},}\ }\href {\doibase
  10.1098/rspa.1927.0118} {\bibfield  {journal} {\bibinfo  {journal}
  {Proceedings of the Royal Society of London. Series A, Containing Papers of a
  Mathematical and Physical Character}\ }\textbf {\bibinfo {volume} {115}},\
  \bibinfo {pages} {700--721} (\bibinfo {year} {1927})}\BibitemShut {NoStop}%
\bibitem [{\citenamefont {Duggan}(2016)}]{SIR1}%
  \BibitemOpen
  \bibfield  {author} {\bibinfo {author} {\bibfnamefont {J.}~\bibnamefont
  {Duggan}},\ }\href@noop {} {\emph {\bibinfo {title} {System Dynamics Modeling
  with R}}}\ (\bibinfo  {publisher} {Springer International Publishing},\
  \bibinfo {address} {Switzerland},\ \bibinfo {year} {2016})\BibitemShut
  {NoStop}%
\bibitem [{\citenamefont {Turner}(2010)}]{SIR2}%
  \BibitemOpen
  \bibfield  {author} {\bibinfo {author} {\bibfnamefont {K.}~\bibnamefont
  {Turner}},\ }\bibfield  {title} {\enquote {\bibinfo {title} {Introduction to
  infectious disease modelling},}\ }\href {\doibase 10.1136/sti.2010.046342}
  {\bibfield  {journal} {\bibinfo  {journal} {Sexually transmitted infections}\
  }\textbf {\bibinfo {volume} {87}} (\bibinfo {year} {2010}),\
  10.1136/sti.2010.046342}\BibitemShut {NoStop}%
\bibitem [{\citenamefont {O'leary}(2004)}]{SIR3}%
  \BibitemOpen
  \bibfield  {author} {\bibinfo {author} {\bibfnamefont {D.}~\bibnamefont
  {O'leary}},\ }\bibfield  {title} {\enquote {\bibinfo {title} {More models of
  infection: It's epidemic},}\ }\href {\doibase 10.1109/MCISE.2004.1289312}
  {\bibfield  {journal} {\bibinfo  {journal} {Computing in Science and
  Engineering}\ }\textbf {\bibinfo {volume} {6}},\ \bibinfo {pages} {70-- 72}
  (\bibinfo {year} {2004})}\BibitemShut {NoStop}%
\bibitem [{\citenamefont {Berger}\ \emph {et~al.}(2020)\citenamefont {Berger},
  \citenamefont {Herkenhoff},\ and\ \citenamefont {Mongey}}]{berger}%
  \BibitemOpen
  \bibfield  {author} {\bibinfo {author} {\bibfnamefont {D.~W.}\ \bibnamefont
  {Berger}}, \bibinfo {author} {\bibfnamefont {K.~F.}\ \bibnamefont
  {Herkenhoff}}, \ and\ \bibinfo {author} {\bibfnamefont {S.}~\bibnamefont
  {Mongey}},\ }\bibfield  {title} {\enquote {\bibinfo {title} {An seir
  infectious disease model with testing and conditional quarantine},}\
  }\href@noop {} {\  (\bibinfo {year} {2020})}\BibitemShut {NoStop}%
\bibitem [{\citenamefont {Chayu~Yang}(2020)}]{yang}%
  \BibitemOpen
  \bibfield  {author} {\bibinfo {author} {\bibfnamefont {J.~W.}\ \bibnamefont
  {Chayu~Yang}},\ }\bibfield  {title} {\enquote {\bibinfo {title} {A
  mathematical model for the novel coronavirus epidemic in wuhan, china},}\
  }\href {\doibase http://dx.doi.org/10.3934/mbe.2020148} {\bibfield  {journal}
  {\bibinfo  {journal} {Mathematical Biosciences and Engineering}\ }\textbf
  {\bibinfo {volume} {17}},\ \bibinfo {pages} {2708} (\bibinfo {year}
  {2020})}\BibitemShut {NoStop}%
\bibitem [{\citenamefont {Grant}(2020)}]{ala}%
  \BibitemOpen
  \bibfield  {author} {\bibinfo {author} {\bibfnamefont {A.}~\bibnamefont
  {Grant}},\ }\bibfield  {title} {\enquote {\bibinfo {title} {Dynamics of
  covid-19 epidemics: Seir models underestimate peak infection rates and
  overestimate epidemic duration},}\ }\href {\doibase
  10.1101/2020.04.02.20050674} {\bibfield  {journal} {\bibinfo  {journal}
  {10.1101/2020.04.02.20050674/medRxiv}\ } (\bibinfo {year} {2020}),\
  10.1101/2020.04.02.20050674}\BibitemShut {NoStop}%
\bibitem [{\citenamefont {Fodor}\ \emph {et~al.}(2020)\citenamefont {Fodor},
  \citenamefont {Katz},\ and\ \citenamefont {Kovacs}}]{fodor}%
  \BibitemOpen
  \bibfield  {author} {\bibinfo {author} {\bibfnamefont {Z.}~\bibnamefont
  {Fodor}}, \bibinfo {author} {\bibfnamefont {S.~D.}\ \bibnamefont {Katz}}, \
  and\ \bibinfo {author} {\bibfnamefont {T.~G.}\ \bibnamefont {Kovacs}},\
  }\href@noop {} {\enquote {\bibinfo {title} {Why integral equations should be
  used instead of differential equations to describe the dynamics of
  epidemics},}\ } (\bibinfo {year} {2020}),\ \Eprint
  {http://arxiv.org/abs/2004.07208} {arXiv:2004.07208 [q-bio.PE]} \BibitemShut
  {NoStop}%
\bibitem [{\citenamefont {Hethcote}(2000)}]{heth}%
  \BibitemOpen
  \bibfield  {author} {\bibinfo {author} {\bibfnamefont {H.~W.}\ \bibnamefont
  {Hethcote}},\ }\bibfield  {title} {\enquote {\bibinfo {title} {The
  mathematics of infectious diseases},}\ }\href {\doibase
  10.1137/S0036144500371907} {\bibfield  {journal} {\bibinfo  {journal} {SIAM
  Review}\ }\textbf {\bibinfo {volume} {42}},\ \bibinfo {pages} {599--653}
  (\bibinfo {year} {2000})}\BibitemShut {NoStop}%
\bibitem [{\citenamefont {Watts}(2004)}]{small}%
  \BibitemOpen
  \bibfield  {author} {\bibinfo {author} {\bibfnamefont {D.~J.}\ \bibnamefont
  {Watts}},\ }\href@noop {} {\emph {\bibinfo {title} {Small worlds: the
  dynamics of networks between order and randomness}}},\ Vol.~\bibinfo {volume}
  {9}\ (\bibinfo  {publisher} {Princeton university press},\ \bibinfo {address}
  {Princeton},\ \bibinfo {year} {2004})\BibitemShut {NoStop}%
\bibitem [{\citenamefont {Colizza}\ \emph {et~al.}(2007)\citenamefont
  {Colizza}, \citenamefont {Pastor-Satorras},\ and\ \citenamefont
  {Vespignani}}]{meta0}%
  \BibitemOpen
  \bibfield  {author} {\bibinfo {author} {\bibfnamefont {V.}~\bibnamefont
  {Colizza}}, \bibinfo {author} {\bibfnamefont {R.}~\bibnamefont
  {Pastor-Satorras}}, \ and\ \bibinfo {author} {\bibfnamefont {A.}~\bibnamefont
  {Vespignani}},\ }\bibfield  {title} {\enquote {\bibinfo {title}
  {Reaction--diffusion processes and metapopulation models in heterogeneous
  networks},}\ }\href {\doibase 10.1038/nphys560} {\bibfield  {journal}
  {\bibinfo  {journal} {Nature Physics}\ }\textbf {\bibinfo {volume} {3}},\
  \bibinfo {pages} {276--282} (\bibinfo {year} {2007})}\BibitemShut {NoStop}%
\bibitem [{\citenamefont {Soriano-Pa{\~{n}}os}\ \emph
  {et~al.}(2018)\citenamefont {Soriano-Pa{\~{n}}os}, \citenamefont
  {Arias-Castro}, \citenamefont {Naranjo-Mayorga},\ and\ \citenamefont
  {G{\'o}mez-Garde{\~{n}}es}}]{Soriano}%
  \BibitemOpen
  \bibfield  {author} {\bibinfo {author} {\bibfnamefont {D.}~\bibnamefont
  {Soriano-Pa{\~{n}}os}}, \bibinfo {author} {\bibfnamefont {H.}~\bibnamefont
  {Arias-Castro}}, \bibinfo {author} {\bibfnamefont {F.}~\bibnamefont
  {Naranjo-Mayorga}}, \ and\ \bibinfo {author} {\bibfnamefont {J.}~\bibnamefont
  {G{\'o}mez-Garde{\~{n}}es}},\ }\bibfield  {title} {\enquote {\bibinfo {title}
  {Impact of human-human contagions in the spread of vector-borne diseases},}\
  }\href {\doibase 10.1140/epjst/e2018-00099-3} {\bibfield  {journal} {\bibinfo
   {journal} {The European Physical Journal Special Topics}\ }\textbf {\bibinfo
  {volume} {227}},\ \bibinfo {pages} {661--672} (\bibinfo {year}
  {2018})}\BibitemShut {NoStop}%
\bibitem [{\citenamefont {Pastor-Satorras}\ \emph {et~al.}(2015)\citenamefont
  {Pastor-Satorras}, \citenamefont {Castellano}, \citenamefont {Van~Mieghem},\
  and\ \citenamefont {Vespignani}}]{meta1}%
  \BibitemOpen
  \bibfield  {author} {\bibinfo {author} {\bibfnamefont {R.}~\bibnamefont
  {Pastor-Satorras}}, \bibinfo {author} {\bibfnamefont {C.}~\bibnamefont
  {Castellano}}, \bibinfo {author} {\bibfnamefont {P.}~\bibnamefont
  {Van~Mieghem}}, \ and\ \bibinfo {author} {\bibfnamefont {A.}~\bibnamefont
  {Vespignani}},\ }\bibfield  {title} {\enquote {\bibinfo {title} {Epidemic
  processes in complex networks},}\ }\href {\doibase 10.1103/RevModPhys.87.925}
  {\bibfield  {journal} {\bibinfo  {journal} {Rev. Mod. Phys.}\ }\textbf
  {\bibinfo {volume} {87}},\ \bibinfo {pages} {925--979} (\bibinfo {year}
  {2015})}\BibitemShut {NoStop}%
\bibitem [{\citenamefont {Colizza}\ and\ \citenamefont
  {Vespignani}(2008)}]{meta2}%
  \BibitemOpen
  \bibfield  {author} {\bibinfo {author} {\bibfnamefont {V.}~\bibnamefont
  {Colizza}}\ and\ \bibinfo {author} {\bibfnamefont {A.}~\bibnamefont
  {Vespignani}},\ }\bibfield  {title} {\enquote {\bibinfo {title} {Epidemic
  modeling in metapopulation systems with heterogeneous coupling pattern:
  Theory and simulations},}\ }\href {\doibase
  https://doi.org/10.1016/j.jtbi.2007.11.028} {\bibfield  {journal} {\bibinfo
  {journal} {Journal of Theoretical Biology}\ }\textbf {\bibinfo {volume}
  {251}},\ \bibinfo {pages} {450 -- 467} (\bibinfo {year} {2008})}\BibitemShut
  {NoStop}%
\bibitem [{\citenamefont {Colizza}\ and\ \citenamefont
  {Vespignani}(2007)}]{meta3}%
  \BibitemOpen
  \bibfield  {author} {\bibinfo {author} {\bibfnamefont {V.}~\bibnamefont
  {Colizza}}\ and\ \bibinfo {author} {\bibfnamefont {A.}~\bibnamefont
  {Vespignani}},\ }\bibfield  {title} {\enquote {\bibinfo {title} {Invasion
  threshold in heterogeneous metapopulation networks},}\ }\href {\doibase
  10.1103/PhysRevLett.99.148701} {\bibfield  {journal} {\bibinfo  {journal}
  {Phys. Rev. Lett.}\ }\textbf {\bibinfo {volume} {99}},\ \bibinfo {pages}
  {148701} (\bibinfo {year} {2007})}\BibitemShut {NoStop}%
\bibitem [{\citenamefont {Belik}\ \emph {et~al.}(2011)\citenamefont {Belik},
  \citenamefont {Geisel},\ and\ \citenamefont {Brockmann}}]{meta4}%
  \BibitemOpen
  \bibfield  {author} {\bibinfo {author} {\bibfnamefont {V.}~\bibnamefont
  {Belik}}, \bibinfo {author} {\bibfnamefont {T.}~\bibnamefont {Geisel}}, \
  and\ \bibinfo {author} {\bibfnamefont {D.}~\bibnamefont {Brockmann}},\
  }\bibfield  {title} {\enquote {\bibinfo {title} {Natural human mobility
  patterns and spatial spread of infectious diseases},}\ }\href {\doibase
  10.1103/PhysRevX.1.011001} {\bibfield  {journal} {\bibinfo  {journal} {Phys.
  Rev. X}\ }\textbf {\bibinfo {volume} {1}},\ \bibinfo {pages} {011001}
  (\bibinfo {year} {2011})}\BibitemShut {NoStop}%
\bibitem [{\citenamefont {de~Arruda}\ \emph
  {et~al.}(2020{\natexlab{a}})\citenamefont {de~Arruda}, \citenamefont
  {Petri},\ and\ \citenamefont {Moreno}}]{network11}%
  \BibitemOpen
  \bibfield  {author} {\bibinfo {author} {\bibfnamefont {G.~F.}\ \bibnamefont
  {de~Arruda}}, \bibinfo {author} {\bibfnamefont {G.}~\bibnamefont {Petri}}, \
  and\ \bibinfo {author} {\bibfnamefont {Y.}~\bibnamefont {Moreno}},\
  }\bibfield  {title} {\enquote {\bibinfo {title} {Social contagion models on
  hypergraphs},}\ }\href {\doibase 10.1103/PhysRevResearch.2.023032} {\bibfield
   {journal} {\bibinfo  {journal} {Phys. Rev. Research}\ }\textbf {\bibinfo
  {volume} {2}},\ \bibinfo {pages} {023032} (\bibinfo {year}
  {2020}{\natexlab{a}})}\BibitemShut {NoStop}%
\bibitem [{\citenamefont {Luo}\ and\ \citenamefont
  {Schaposnik}(2020)}]{network10}%
  \BibitemOpen
  \bibfield  {author} {\bibinfo {author} {\bibfnamefont {Y.}~\bibnamefont
  {Luo}}\ and\ \bibinfo {author} {\bibfnamefont {L.~P.}\ \bibnamefont
  {Schaposnik}},\ }\bibfield  {title} {\enquote {\bibinfo {title} {Minimal
  percolating sets for mutating infectious diseases},}\ }\href {\doibase
  10.1103/PhysRevResearch.2.023001} {\bibfield  {journal} {\bibinfo  {journal}
  {Phys. Rev. Research}\ }\textbf {\bibinfo {volume} {2}},\ \bibinfo {pages}
  {023001} (\bibinfo {year} {2020})}\BibitemShut {NoStop}%
\bibitem [{\citenamefont {Chen}\ \emph {et~al.}(2020)\citenamefont {Chen},
  \citenamefont {Hu},\ and\ \citenamefont {Li}}]{network6}%
  \BibitemOpen
  \bibfield  {author} {\bibinfo {author} {\bibfnamefont {J.}~\bibnamefont
  {Chen}}, \bibinfo {author} {\bibfnamefont {M.-B.}\ \bibnamefont {Hu}}, \ and\
  \bibinfo {author} {\bibfnamefont {M.}~\bibnamefont {Li}},\ }\bibfield
  {title} {\enquote {\bibinfo {title} {Traffic-driven epidemic spreading in
  multiplex networks},}\ }\href {\doibase 10.1103/PhysRevE.101.012301}
  {\bibfield  {journal} {\bibinfo  {journal} {Phys. Rev. E}\ }\textbf {\bibinfo
  {volume} {101}},\ \bibinfo {pages} {012301} (\bibinfo {year}
  {2020})}\BibitemShut {NoStop}%
\bibitem [{\citenamefont {Van~Mieghem}\ and\ \citenamefont
  {Liu}(2019)}]{network2}%
  \BibitemOpen
  \bibfield  {author} {\bibinfo {author} {\bibfnamefont {P.}~\bibnamefont
  {Van~Mieghem}}\ and\ \bibinfo {author} {\bibfnamefont {Q.}~\bibnamefont
  {Liu}},\ }\bibfield  {title} {\enquote {\bibinfo {title} {Explicit
  non-markovian susceptible-infected-susceptible mean-field epidemic threshold
  for weibull and gamma infections but poisson curings},}\ }\href {\doibase
  10.1103/PhysRevE.100.022317} {\bibfield  {journal} {\bibinfo  {journal}
  {Phys. Rev. E}\ }\textbf {\bibinfo {volume} {100}},\ \bibinfo {pages}
  {022317} (\bibinfo {year} {2019})}\BibitemShut {NoStop}%
\bibitem [{\citenamefont {de~Arruda}\ \emph
  {et~al.}(2020{\natexlab{b}})\citenamefont {de~Arruda}, \citenamefont {Petri},
  \citenamefont {Rodrigues},\ and\ \citenamefont {Moreno}}]{network7}%
  \BibitemOpen
  \bibfield  {author} {\bibinfo {author} {\bibfnamefont {G.~F.}\ \bibnamefont
  {de~Arruda}}, \bibinfo {author} {\bibfnamefont {G.}~\bibnamefont {Petri}},
  \bibinfo {author} {\bibfnamefont {F.~A.}\ \bibnamefont {Rodrigues}}, \ and\
  \bibinfo {author} {\bibfnamefont {Y.}~\bibnamefont {Moreno}},\ }\bibfield
  {title} {\enquote {\bibinfo {title} {Impact of the distribution of recovery
  rates on disease spreading in complex networks},}\ }\href {\doibase
  10.1103/PhysRevResearch.2.013046} {\bibfield  {journal} {\bibinfo  {journal}
  {Phys. Rev. Research}\ }\textbf {\bibinfo {volume} {2}},\ \bibinfo {pages}
  {013046} (\bibinfo {year} {2020}{\natexlab{b}})}\BibitemShut {NoStop}%
\bibitem [{\citenamefont {Castellano}\ and\ \citenamefont
  {Pastor-Satorras}(2020)}]{network9}%
  \BibitemOpen
  \bibfield  {author} {\bibinfo {author} {\bibfnamefont {C.}~\bibnamefont
  {Castellano}}\ and\ \bibinfo {author} {\bibfnamefont {R.}~\bibnamefont
  {Pastor-Satorras}},\ }\bibfield  {title} {\enquote {\bibinfo {title}
  {Cumulative merging percolation and the epidemic transition of the
  susceptible-infected-susceptible model in networks},}\ }\href {\doibase
  10.1103/PhysRevX.10.011070} {\bibfield  {journal} {\bibinfo  {journal} {Phys.
  Rev. X}\ }\textbf {\bibinfo {volume} {10}},\ \bibinfo {pages} {011070}
  (\bibinfo {year} {2020})}\BibitemShut {NoStop}%
\bibitem [{\citenamefont {Moore}\ and\ \citenamefont
  {Rogers}(2020)}]{network8}%
  \BibitemOpen
  \bibfield  {author} {\bibinfo {author} {\bibfnamefont {S.}~\bibnamefont
  {Moore}}\ and\ \bibinfo {author} {\bibfnamefont {T.}~\bibnamefont {Rogers}},\
  }\bibfield  {title} {\enquote {\bibinfo {title} {Predicting the speed of
  epidemics spreading in networks},}\ }\href {\doibase
  10.1103/PhysRevLett.124.068301} {\bibfield  {journal} {\bibinfo  {journal}
  {Phys. Rev. Lett.}\ }\textbf {\bibinfo {volume} {124}},\ \bibinfo {pages}
  {068301} (\bibinfo {year} {2020})}\BibitemShut {NoStop}%
\bibitem [{\citenamefont {Lee}\ \emph {et~al.}(2019)\citenamefont {Lee},
  \citenamefont {Emmons}, \citenamefont {Gibson}, \citenamefont {Moody},\ and\
  \citenamefont {Mucha}}]{network5}%
  \BibitemOpen
  \bibfield  {author} {\bibinfo {author} {\bibfnamefont {E.}~\bibnamefont
  {Lee}}, \bibinfo {author} {\bibfnamefont {S.}~\bibnamefont {Emmons}},
  \bibinfo {author} {\bibfnamefont {R.}~\bibnamefont {Gibson}}, \bibinfo
  {author} {\bibfnamefont {J.}~\bibnamefont {Moody}}, \ and\ \bibinfo {author}
  {\bibfnamefont {P.~J.}\ \bibnamefont {Mucha}},\ }\bibfield  {title} {\enquote
  {\bibinfo {title} {Concurrency and reachability in treelike temporal
  networks},}\ }\href {\doibase 10.1103/PhysRevE.100.062305} {\bibfield
  {journal} {\bibinfo  {journal} {Phys. Rev. E}\ }\textbf {\bibinfo {volume}
  {100}},\ \bibinfo {pages} {062305} (\bibinfo {year} {2019})}\BibitemShut
  {NoStop}%
\bibitem [{\citenamefont {Silva}\ \emph {et~al.}(2019)\citenamefont {Silva},
  \citenamefont {Ferreira}, \citenamefont {Cota}, \citenamefont
  {Pastor-Satorras},\ and\ \citenamefont {Castellano}}]{network3}%
  \BibitemOpen
  \bibfield  {author} {\bibinfo {author} {\bibfnamefont {D.~H.}\ \bibnamefont
  {Silva}}, \bibinfo {author} {\bibfnamefont {S.~C.}\ \bibnamefont {Ferreira}},
  \bibinfo {author} {\bibfnamefont {W.}~\bibnamefont {Cota}}, \bibinfo {author}
  {\bibfnamefont {R.}~\bibnamefont {Pastor-Satorras}}, \ and\ \bibinfo {author}
  {\bibfnamefont {C.}~\bibnamefont {Castellano}},\ }\bibfield  {title}
  {\enquote {\bibinfo {title} {Spectral properties and the accuracy of
  mean-field approaches for epidemics on correlated power-law networks},}\
  }\href {\doibase 10.1103/PhysRevResearch.1.033024} {\bibfield  {journal}
  {\bibinfo  {journal} {Phys. Rev. Research}\ }\textbf {\bibinfo {volume}
  {1}},\ \bibinfo {pages} {033024} (\bibinfo {year} {2019})}\BibitemShut
  {NoStop}%
\bibitem [{\citenamefont {Bonamassa}\ \emph {et~al.}(2019)\citenamefont
  {Bonamassa}, \citenamefont {Gross}, \citenamefont {Danziger},\ and\
  \citenamefont {Havlin}}]{network1}%
  \BibitemOpen
  \bibfield  {author} {\bibinfo {author} {\bibfnamefont {I.}~\bibnamefont
  {Bonamassa}}, \bibinfo {author} {\bibfnamefont {B.}~\bibnamefont {Gross}},
  \bibinfo {author} {\bibfnamefont {M.~M.}\ \bibnamefont {Danziger}}, \ and\
  \bibinfo {author} {\bibfnamefont {S.}~\bibnamefont {Havlin}},\ }\bibfield
  {title} {\enquote {\bibinfo {title} {Critical stretching of mean-field
  regimes in spatial networks},}\ }\href {\doibase
  10.1103/PhysRevLett.123.088301} {\bibfield  {journal} {\bibinfo  {journal}
  {Phys. Rev. Lett.}\ }\textbf {\bibinfo {volume} {123}},\ \bibinfo {pages}
  {088301} (\bibinfo {year} {2019})}\BibitemShut {NoStop}%
\bibitem [{\citenamefont {Mello}(2020)}]{one-way}%
  \BibitemOpen
  \bibfield  {author} {\bibinfo {author} {\bibfnamefont {B.~A.}\ \bibnamefont
  {Mello}},\ }\bibfield  {title} {\enquote {\bibinfo {title} {One-way
  pedestrian traffic is a means of reducing personal encounters in
  epidemics},}\ }\href@noop {} {\  (\bibinfo {year} {2020})},\ \Eprint
  {http://arxiv.org/abs/2004.00423} {arXiv:2004.00423 [physics.soc-ph]}
  \BibitemShut {NoStop}%
\bibitem [{\citenamefont {Weng}\ \emph {et~al.}(2006)\citenamefont {Weng},
  \citenamefont {Chen}, \citenamefont {Yuan},\ and\ \citenamefont {Fan}}]{CA1}%
  \BibitemOpen
  \bibfield  {author} {\bibinfo {author} {\bibfnamefont {W.}~\bibnamefont
  {Weng}}, \bibinfo {author} {\bibfnamefont {T.}~\bibnamefont {Chen}}, \bibinfo
  {author} {\bibfnamefont {H.}~\bibnamefont {Yuan}}, \ and\ \bibinfo {author}
  {\bibfnamefont {W.}~\bibnamefont {Fan}},\ }\bibfield  {title} {\enquote
  {\bibinfo {title} {Cellular automaton simulation of pedestrian counter flow
  with different walk velocities},}\ }\href {\doibase
  10.1103/PHYSREVE.74.036102} {\bibfield  {journal} {\bibinfo  {journal}
  {Physical review. E, Statistical, nonlinear, and soft matter physics}\
  }\textbf {\bibinfo {volume} {74}},\ \bibinfo {pages} {036102} (\bibinfo
  {year} {2006})}\BibitemShut {NoStop}%
\bibitem [{\citenamefont {Feng}\ \emph {et~al.}(2013)\citenamefont {Feng},
  \citenamefont {Ding}, \citenamefont {Chen},\ and\ \citenamefont
  {Zhang}}]{CA2}%
  \BibitemOpen
  \bibfield  {author} {\bibinfo {author} {\bibfnamefont {S.}~\bibnamefont
  {Feng}}, \bibinfo {author} {\bibfnamefont {N.}~\bibnamefont {Ding}}, \bibinfo
  {author} {\bibfnamefont {T.}~\bibnamefont {Chen}}, \ and\ \bibinfo {author}
  {\bibfnamefont {H.}~\bibnamefont {Zhang}},\ }\bibfield  {title} {\enquote
  {\bibinfo {title} {Simulation of pedestrian flow based on cellular automata:
  A case of pedestrian crossing street at section in china},}\ }\href {\doibase
  10.1016/j.physa.2013.03.008} {\bibfield  {journal} {\bibinfo  {journal}
  {Physica A: Statistical Mechanics and its Applications}\ }\textbf {\bibinfo
  {volume} {392}},\ \bibinfo {pages} {2847–2859} (\bibinfo {year}
  {2013})}\BibitemShut {NoStop}%
\bibitem [{\citenamefont {Liu}\ \emph {et~al.}(2017)\citenamefont {Liu},
  \citenamefont {Zeng}, \citenamefont {Chen},\ and\ \citenamefont {Wu}}]{CA3}%
  \BibitemOpen
  \bibfield  {author} {\bibinfo {author} {\bibfnamefont {M.}~\bibnamefont
  {Liu}}, \bibinfo {author} {\bibfnamefont {W.}~\bibnamefont {Zeng}}, \bibinfo
  {author} {\bibfnamefont {P.}~\bibnamefont {Chen}}, \ and\ \bibinfo {author}
  {\bibfnamefont {X.}~\bibnamefont {Wu}},\ }\bibfield  {title} {\enquote
  {\bibinfo {title} {A microscopic simulation model for pedestrian-pedestrian
  and pedestrian-vehicle interactions at crosswalks},}\ }\href {\doibase
  10.1371/journal.pone.0180992} {\bibfield  {journal} {\bibinfo  {journal}
  {PLoS ONE}\ }\textbf {\bibinfo {volume} {12}} (\bibinfo {year} {2017}),\
  10.1371/journal.pone.0180992}\BibitemShut {NoStop}%
\bibitem [{\citenamefont {Jiang}\ \emph {et~al.}(2020)\citenamefont {Jiang},
  \citenamefont {Rayner},\ and\ \citenamefont {Luo}}]{Covid1}%
  \BibitemOpen
  \bibfield  {author} {\bibinfo {author} {\bibfnamefont {X.}~\bibnamefont
  {Jiang}}, \bibinfo {author} {\bibfnamefont {S.}~\bibnamefont {Rayner}}, \
  and\ \bibinfo {author} {\bibfnamefont {M.-H.}\ \bibnamefont {Luo}},\
  }\bibfield  {title} {\enquote {\bibinfo {title} {Does sars‐cov‐2 has a
  longer incubation period than sars and mers?}}\ }\href {\doibase
  10.1002/jmv.25708} {\bibfield  {journal} {\bibinfo  {journal} {Journal of
  Medical Virology}\ }\textbf {\bibinfo {volume} {92}} (\bibinfo {year}
  {2020}),\ 10.1002/jmv.25708}\BibitemShut {NoStop}%
\bibitem [{\citenamefont {Liu}\ \emph {et~al.}(2020)\citenamefont {Liu},
  \citenamefont {Gayle}, \citenamefont {Wilder-Smith},\ and\ \citenamefont
  {Rocklöv}}]{Covid2}%
  \BibitemOpen
  \bibfield  {author} {\bibinfo {author} {\bibfnamefont {Y.}~\bibnamefont
  {Liu}}, \bibinfo {author} {\bibfnamefont {A.}~\bibnamefont {Gayle}}, \bibinfo
  {author} {\bibfnamefont {A.}~\bibnamefont {Wilder-Smith}}, \ and\ \bibinfo
  {author} {\bibfnamefont {J.}~\bibnamefont {Rocklöv}},\ }\bibfield  {title}
  {\enquote {\bibinfo {title} {The reproductive number of covid-19 is higher
  compared to sars coronavirus},}\ }\href {\doibase 10.1093/jtm/taaa021}
  {\bibfield  {journal} {\bibinfo  {journal} {Journal of travel medicine}\
  }\textbf {\bibinfo {volume} {27}} (\bibinfo {year} {2020}),\
  10.1093/jtm/taaa021}\BibitemShut {NoStop}%
\bibitem [{\citenamefont {Kucharski}\ \emph {et~al.}(2020)\citenamefont
  {Kucharski}, \citenamefont {Russell}, \citenamefont {Diamond}, \citenamefont
  {Liu}, \citenamefont {Edmunds}, \citenamefont {Funk}, \citenamefont {Eggo},
  \citenamefont {Sun}, \citenamefont {Jit}, \citenamefont {Munday},
  \citenamefont {Davies}, \citenamefont {Gimma}, \citenamefont {Zandvoort},
  \citenamefont {Gibbs}, \citenamefont {Hellewell}, \citenamefont {Jarvis},
  \citenamefont {Clifford}, \citenamefont {Quilty}, \citenamefont {Bosse},\
  and\ \citenamefont {Flasche}}]{Covid3}%
  \BibitemOpen
  \bibfield  {author} {\bibinfo {author} {\bibfnamefont {A.}~\bibnamefont
  {Kucharski}}, \bibinfo {author} {\bibfnamefont {T.}~\bibnamefont {Russell}},
  \bibinfo {author} {\bibfnamefont {C.}~\bibnamefont {Diamond}}, \bibinfo
  {author} {\bibfnamefont {Y.}~\bibnamefont {Liu}}, \bibinfo {author}
  {\bibfnamefont {J.}~\bibnamefont {Edmunds}}, \bibinfo {author} {\bibfnamefont
  {S.}~\bibnamefont {Funk}}, \bibinfo {author} {\bibfnamefont {R.}~\bibnamefont
  {Eggo}}, \bibinfo {author} {\bibfnamefont {F.}~\bibnamefont {Sun}}, \bibinfo
  {author} {\bibfnamefont {M.}~\bibnamefont {Jit}}, \bibinfo {author}
  {\bibfnamefont {J.}~\bibnamefont {Munday}}, \bibinfo {author} {\bibfnamefont
  {N.}~\bibnamefont {Davies}}, \bibinfo {author} {\bibfnamefont
  {A.}~\bibnamefont {Gimma}}, \bibinfo {author} {\bibfnamefont
  {K.}~\bibnamefont {Zandvoort}}, \bibinfo {author} {\bibfnamefont
  {H.}~\bibnamefont {Gibbs}}, \bibinfo {author} {\bibfnamefont
  {J.}~\bibnamefont {Hellewell}}, \bibinfo {author} {\bibfnamefont
  {C.}~\bibnamefont {Jarvis}}, \bibinfo {author} {\bibfnamefont
  {S.}~\bibnamefont {Clifford}}, \bibinfo {author} {\bibfnamefont
  {B.}~\bibnamefont {Quilty}}, \bibinfo {author} {\bibfnamefont
  {N.}~\bibnamefont {Bosse}}, \ and\ \bibinfo {author} {\bibfnamefont
  {S.}~\bibnamefont {Flasche}},\ }\bibfield  {title} {\enquote {\bibinfo
  {title} {Early dynamics of transmission and control of covid-19: a
  mathematical modelling study},}\ }\href {\doibase
  10.1016/S1473-3099(20)30144-4} {\bibfield  {journal} {\bibinfo  {journal}
  {The Lancet Infectious Diseases}\ }\textbf {\bibinfo {volume} {20}} (\bibinfo
  {year} {2020}),\ 10.1016/S1473-3099(20)30144-4}\BibitemShut {NoStop}%
\bibitem [{\citenamefont {Ferretti}\ \emph {et~al.}(2020)\citenamefont
  {Ferretti}, \citenamefont {Wymant}, \citenamefont {Kendall}, \citenamefont
  {Zhao}, \citenamefont {Nurtay}, \citenamefont {Abeler-Dörner}, \citenamefont
  {Parker}, \citenamefont {Bonsall},\ and\ \citenamefont {Fraser}}]{Covid4}%
  \BibitemOpen
  \bibfield  {author} {\bibinfo {author} {\bibfnamefont {L.}~\bibnamefont
  {Ferretti}}, \bibinfo {author} {\bibfnamefont {C.}~\bibnamefont {Wymant}},
  \bibinfo {author} {\bibfnamefont {M.}~\bibnamefont {Kendall}}, \bibinfo
  {author} {\bibfnamefont {L.}~\bibnamefont {Zhao}}, \bibinfo {author}
  {\bibfnamefont {A.}~\bibnamefont {Nurtay}}, \bibinfo {author} {\bibfnamefont
  {L.}~\bibnamefont {Abeler-Dörner}}, \bibinfo {author} {\bibfnamefont
  {M.}~\bibnamefont {Parker}}, \bibinfo {author} {\bibfnamefont
  {D.}~\bibnamefont {Bonsall}}, \ and\ \bibinfo {author} {\bibfnamefont
  {C.}~\bibnamefont {Fraser}},\ }\bibfield  {title} {\enquote {\bibinfo {title}
  {Quantifying sars-cov-2 transmission suggests epidemic control with digital
  contact tracing},}\ }\href {\doibase 10.1126/science.abb6936} {\bibfield
  {journal} {\bibinfo  {journal} {Science}\ }\textbf {\bibinfo {volume}
  {368}},\ \bibinfo {pages} {eabb6936} (\bibinfo {year} {2020})}\BibitemShut
  {NoStop}%
\bibitem [{\citenamefont {Maerivoet}\ and\ \citenamefont {{De
  Moor}}(2005)}]{cellulartraffic}%
  \BibitemOpen
  \bibfield  {author} {\bibinfo {author} {\bibfnamefont {S.}~\bibnamefont
  {Maerivoet}}\ and\ \bibinfo {author} {\bibfnamefont {B.}~\bibnamefont {{De
  Moor}}},\ }\bibfield  {title} {\enquote {\bibinfo {title} {Cellular automata
  models of road traffic},}\ }\href {\doibase
  https://doi.org/10.1016/j.physrep.2005.08.005} {\bibfield  {journal}
  {\bibinfo  {journal} {Physics Reports}\ }\textbf {\bibinfo {volume} {419}},\
  \bibinfo {pages} {1 -- 64} (\bibinfo {year} {2005})}\BibitemShut {NoStop}%
\bibitem [{\citenamefont {Chopard}(2012)}]{CALB}%
  \BibitemOpen
  \bibfield  {author} {\bibinfo {author} {\bibfnamefont {B.}~\bibnamefont
  {Chopard}},\ }\enquote {\bibinfo {title} {Cellular automata and lattice
  boltzmann modeling of physical systems},}\ in\ \href {\doibase
  10.1007/978-3-540-92910-9_9} {\emph {\bibinfo {booktitle} {Handbook of
  Natural Computing}}},\ \bibinfo {editor} {edited by\ \bibinfo {editor}
  {\bibfnamefont {G.}~\bibnamefont {Rozenberg}}, \bibinfo {editor}
  {\bibfnamefont {T.}~\bibnamefont {B{\"a}ck}}, \ and\ \bibinfo {editor}
  {\bibfnamefont {J.~N.}\ \bibnamefont {Kok}}}\ (\bibinfo  {publisher}
  {Springer Berlin Heidelberg},\ \bibinfo {address} {Berlin, Heidelberg},\
  \bibinfo {year} {2012})\ pp.\ \bibinfo {pages} {287--331}\BibitemShut
  {NoStop}%
\bibitem [{\citenamefont {Chen}\ and\ \citenamefont
  {Doolen}(1998)}]{latticeboltzmann}%
  \BibitemOpen
  \bibfield  {author} {\bibinfo {author} {\bibfnamefont {S.}~\bibnamefont
  {Chen}}\ and\ \bibinfo {author} {\bibfnamefont {G.~D.}\ \bibnamefont
  {Doolen}},\ }\bibfield  {title} {\enquote {\bibinfo {title} {Lattice
  boltzmann method for fluid flows},}\ }\href {\doibase
  10.1146/annurev.fluid.30.1.329} {\bibfield  {journal} {\bibinfo  {journal}
  {Annual Review of Fluid Mechanics}\ }\textbf {\bibinfo {volume} {30}},\
  \bibinfo {pages} {329--364} (\bibinfo {year} {1998})}\BibitemShut {NoStop}%
\bibitem [{\citenamefont {Guan}\ \emph {et~al.}(2020)\citenamefont {Guan},
  \citenamefont {Ni}, \citenamefont {Hu}, \citenamefont {Liang}, \citenamefont
  {Ou}, \citenamefont {He}, \citenamefont {Liu}, \citenamefont {Shan},
  \citenamefont {Lei}, \citenamefont {Hui}, \citenamefont {Du}, \citenamefont
  {Li}, \citenamefont {Zeng}, \citenamefont {Yuen}, \citenamefont {Chen},
  \citenamefont {Tang}, \citenamefont {Wang}, \citenamefont {Chen},
  \citenamefont {Xiang}, \citenamefont {Li}, \citenamefont {Wang},
  \citenamefont {Liang}, \citenamefont {Peng}, \citenamefont {Wei},
  \citenamefont {Liu}, \citenamefont {Hu}, \citenamefont {Peng}, \citenamefont
  {Wang}, \citenamefont {Liu}, \citenamefont {Chen}, \citenamefont {Li},
  \citenamefont {Zheng}, \citenamefont {Qiu}, \citenamefont {Luo},
  \citenamefont {Ye}, \citenamefont {Zhu},\ and\ \citenamefont
  {Zhong}}]{Covid5}%
  \BibitemOpen
  \bibfield  {author} {\bibinfo {author} {\bibfnamefont {W.-j.}\ \bibnamefont
  {Guan}}, \bibinfo {author} {\bibfnamefont {Z.-y.}\ \bibnamefont {Ni}},
  \bibinfo {author} {\bibfnamefont {Y.}~\bibnamefont {Hu}}, \bibinfo {author}
  {\bibfnamefont {W.-h.}\ \bibnamefont {Liang}}, \bibinfo {author}
  {\bibfnamefont {C.-q.}\ \bibnamefont {Ou}}, \bibinfo {author} {\bibfnamefont
  {J.-x.}\ \bibnamefont {He}}, \bibinfo {author} {\bibfnamefont
  {L.}~\bibnamefont {Liu}}, \bibinfo {author} {\bibfnamefont {H.}~\bibnamefont
  {Shan}}, \bibinfo {author} {\bibfnamefont {C.-l.}\ \bibnamefont {Lei}},
  \bibinfo {author} {\bibfnamefont {D.~S.}\ \bibnamefont {Hui}}, \bibinfo
  {author} {\bibfnamefont {B.}~\bibnamefont {Du}}, \bibinfo {author}
  {\bibfnamefont {L.-j.}\ \bibnamefont {Li}}, \bibinfo {author} {\bibfnamefont
  {G.}~\bibnamefont {Zeng}}, \bibinfo {author} {\bibfnamefont {K.-Y.}\
  \bibnamefont {Yuen}}, \bibinfo {author} {\bibfnamefont {R.-c.}\ \bibnamefont
  {Chen}}, \bibinfo {author} {\bibfnamefont {C.-l.}\ \bibnamefont {Tang}},
  \bibinfo {author} {\bibfnamefont {T.}~\bibnamefont {Wang}}, \bibinfo {author}
  {\bibfnamefont {P.-y.}\ \bibnamefont {Chen}}, \bibinfo {author}
  {\bibfnamefont {J.}~\bibnamefont {Xiang}}, \bibinfo {author} {\bibfnamefont
  {S.-y.}\ \bibnamefont {Li}}, \bibinfo {author} {\bibfnamefont {J.-l.}\
  \bibnamefont {Wang}}, \bibinfo {author} {\bibfnamefont {Z.-j.}\ \bibnamefont
  {Liang}}, \bibinfo {author} {\bibfnamefont {Y.-x.}\ \bibnamefont {Peng}},
  \bibinfo {author} {\bibfnamefont {L.}~\bibnamefont {Wei}}, \bibinfo {author}
  {\bibfnamefont {Y.}~\bibnamefont {Liu}}, \bibinfo {author} {\bibfnamefont
  {Y.-h.}\ \bibnamefont {Hu}}, \bibinfo {author} {\bibfnamefont
  {P.}~\bibnamefont {Peng}}, \bibinfo {author} {\bibfnamefont {J.-m.}\
  \bibnamefont {Wang}}, \bibinfo {author} {\bibfnamefont {J.-y.}\ \bibnamefont
  {Liu}}, \bibinfo {author} {\bibfnamefont {Z.}~\bibnamefont {Chen}}, \bibinfo
  {author} {\bibfnamefont {G.}~\bibnamefont {Li}}, \bibinfo {author}
  {\bibfnamefont {Z.-j.}\ \bibnamefont {Zheng}}, \bibinfo {author}
  {\bibfnamefont {S.-q.}\ \bibnamefont {Qiu}}, \bibinfo {author} {\bibfnamefont
  {J.}~\bibnamefont {Luo}}, \bibinfo {author} {\bibfnamefont {C.-j.}\
  \bibnamefont {Ye}}, \bibinfo {author} {\bibfnamefont {S.-y.}\ \bibnamefont
  {Zhu}}, \ and\ \bibinfo {author} {\bibfnamefont {N.-s.}\ \bibnamefont
  {Zhong}},\ }\bibfield  {title} {\enquote {\bibinfo {title} {Clinical
  characteristics of 2019 novel coronavirus infection in china},}\ }\href
  {\doibase 10.1101/2020.02.06.20020974} {\bibfield  {journal} {\bibinfo
  {journal} {medRxiv}\ } (\bibinfo {year} {2020}),\
  10.1101/2020.02.06.20020974}\BibitemShut {NoStop}%
\bibitem [{\citenamefont {Nava}\ \emph {et~al.}(2020)\citenamefont {Nava},
  \citenamefont {Rossi},\ and\ \citenamefont {Giuliano}}]{nrg_lindblad}%
  \BibitemOpen
  \bibfield  {author} {\bibinfo {author} {\bibfnamefont {A.}~\bibnamefont
  {Nava}}, \bibinfo {author} {\bibfnamefont {M.}~\bibnamefont {Rossi}}, \ and\
  \bibinfo {author} {\bibfnamefont {D.}~\bibnamefont {Giuliano}},\ }\href@noop
  {} {\enquote {\bibinfo {title} {Lindblad equation approach to the optimal
  working point in non equilibrium stationary states of an interacting
  electronic one-dimensional system: application to the spinless {H}ubbard
  chain in the clean and in the weakly disordered limit},}\ } (\bibinfo {year}
  {2020}),\ \Eprint {http://arxiv.org/abs/2010.04533} {arXiv:2010.04533
  [cond-mat.str-el]} \BibitemShut {NoStop}%
\bibitem [{\citenamefont {Nava-Sede{\~{n}}o}\ \emph {et~al.}(2017)\citenamefont
  {Nava-Sede{\~{n}}o}, \citenamefont {Hatzikirou}, \citenamefont {Peruani},\
  and\ \citenamefont {Deutsch}}]{fokker}%
  \BibitemOpen
  \bibfield  {author} {\bibinfo {author} {\bibfnamefont {J.~M.}\ \bibnamefont
  {Nava-Sede{\~{n}}o}}, \bibinfo {author} {\bibfnamefont {H.}~\bibnamefont
  {Hatzikirou}}, \bibinfo {author} {\bibfnamefont {F.}~\bibnamefont {Peruani}},
  \ and\ \bibinfo {author} {\bibfnamefont {A.}~\bibnamefont {Deutsch}},\
  }\bibfield  {title} {\enquote {\bibinfo {title} {Extracting cellular
  automaton rules from physical langevin equation models for single and
  collective cell migration},}\ }\href {\doibase 10.1007/s00285-017-1106-9}
  {\bibfield  {journal} {\bibinfo  {journal} {Journal of Mathematical Biology}\
  }\textbf {\bibinfo {volume} {75}},\ \bibinfo {pages} {1075--1100} (\bibinfo
  {year} {2017})}\BibitemShut {NoStop}%
\bibitem [{\citenamefont {Thaler}\ and\ \citenamefont
  {Sunstein}(2012)}]{nudge}%
  \BibitemOpen
  \bibfield  {author} {\bibinfo {author} {\bibfnamefont {R.}~\bibnamefont
  {Thaler}}\ and\ \bibinfo {author} {\bibfnamefont {C.}~\bibnamefont
  {Sunstein}},\ }\href@noop {} {\emph {\bibinfo {title} {Nudge: Improving
  Decisions About Health, Wealth and Happiness}}}\ (\bibinfo  {publisher}
  {Penguin Books Limited},\ \bibinfo {address} {USA},\ \bibinfo {year}
  {2012})\BibitemShut {NoStop}%
\bibitem [{not()}]{note}%
  \BibitemOpen
  \href@noop {} {}\bibinfo {note} {Note that here, as well as in the following
  figures, we use pointplots for the numerical data generated with the CA. We
  do so since, in all the cases we discuss here, there are small-amplitude
  wigglings in the plots that would not allow for a smooth, continuous
  interpolation of the data with a full line. In principle, the wigglings can
  be reduced by increasing the simulation time. However, it would not affect
  the global behavior of the plots. Therefore, for this reason, here and in the
  following we decided to draw pointplots, rather than trying to make an
  interpolation of the numerical points with a full line.}\BibitemShut {Stop}%
\bibitem [{\citenamefont {Erban}\ \emph {et~al.}(2007)\citenamefont {Erban},
  \citenamefont {Chapman}, \citenamefont {Philip},\ and\ \citenamefont
  {Maini}}]{gillespie0}%
  \BibitemOpen
  \bibfield  {author} {\bibinfo {author} {\bibfnamefont {R.}~\bibnamefont
  {Erban}}, \bibinfo {author} {\bibfnamefont {S.~J.}\ \bibnamefont {Chapman}},
  \bibinfo {author} {\bibnamefont {Philip}}, \ and\ \bibinfo {author}
  {\bibfnamefont {K.}~\bibnamefont {Maini}},\ }\href@noop {} {\enquote
  {\bibinfo {title} {A practical guide to stochastic simulations of
  reactiondiffusion processes.}}\ } (\bibinfo {year} {2007})\BibitemShut
  {NoStop}%
\bibitem [{\citenamefont {Vestergaard}\ and\ \citenamefont
  {Génois}(2015)}]{gillespie1}%
  \BibitemOpen
  \bibfield  {author} {\bibinfo {author} {\bibfnamefont {C.~L.}\ \bibnamefont
  {Vestergaard}}\ and\ \bibinfo {author} {\bibfnamefont {M.}~\bibnamefont
  {Génois}},\ }\bibfield  {title} {\enquote {\bibinfo {title} {Temporal
  gillespie algorithm: Fast simulation of contagion processes on time-varying
  networks},}\ }\href {\doibase 10.1371/journal.pcbi.1004579} {\bibfield
  {journal} {\bibinfo  {journal} {PLOS Computational Biology}\ }\textbf
  {\bibinfo {volume} {11}},\ \bibinfo {pages} {1--28} (\bibinfo {year}
  {2015})}\BibitemShut {NoStop}%
\bibitem [{\citenamefont {Delamater}\ \emph {et~al.}(2019)\citenamefont
  {Delamater}, \citenamefont {Street}, \citenamefont {Leslie}, \citenamefont
  {Yang},\ and\ \citenamefont {Jacobsen}}]{R0}%
  \BibitemOpen
  \bibfield  {author} {\bibinfo {author} {\bibfnamefont {P.~L.}\ \bibnamefont
  {Delamater}}, \bibinfo {author} {\bibfnamefont {E.~J.}\ \bibnamefont
  {Street}}, \bibinfo {author} {\bibfnamefont {T.~F.}\ \bibnamefont {Leslie}},
  \bibinfo {author} {\bibfnamefont {Y.~T.}\ \bibnamefont {Yang}}, \ and\
  \bibinfo {author} {\bibfnamefont {K.~H.}\ \bibnamefont {Jacobsen}},\
  }\bibfield  {title} {\enquote {\bibinfo {title} {Complexity of the basic
  reproduction number (r0)},}\ }\href@noop {} {\bibfield  {journal} {\bibinfo
  {journal} {Emerging infectious diseases}\ }\textbf {\bibinfo {volume} {25}},\
  \bibinfo {pages} {1} (\bibinfo {year} {2019})}\BibitemShut {NoStop}%
\bibitem [{\citenamefont {Helbing}\ and\ \citenamefont
  {Moln\'ar}(1995)}]{social_force}%
  \BibitemOpen
  \bibfield  {author} {\bibinfo {author} {\bibfnamefont {D.}~\bibnamefont
  {Helbing}}\ and\ \bibinfo {author} {\bibfnamefont {P.}~\bibnamefont
  {Moln\'ar}},\ }\bibfield  {title} {\enquote {\bibinfo {title} {Social force
  model for pedestrian dynamics},}\ }\href {\doibase 10.1103/PhysRevE.51.4282}
  {\bibfield  {journal} {\bibinfo  {journal} {Phys. Rev. E}\ }\textbf {\bibinfo
  {volume} {51}},\ \bibinfo {pages} {4282--4286} (\bibinfo {year}
  {1995})}\BibitemShut {NoStop}%
\bibitem [{\citenamefont {Kermack}\ and\ \citenamefont
  {McKendrick}(1991{\natexlab{a}})}]{kam_1}%
  \BibitemOpen
  \bibfield  {author} {\bibinfo {author} {\bibfnamefont {W.~O.}\ \bibnamefont
  {Kermack}}\ and\ \bibinfo {author} {\bibfnamefont {A.~G.}\ \bibnamefont
  {McKendrick}},\ }\bibfield  {title} {\enquote {\bibinfo {title}
  {Contributions to the mathematical theory of epidemics i},}\ }\href {\doibase
  10.1007/BF02464423} {\bibfield  {journal} {\bibinfo  {journal} {Bulletin of
  Mathematical Biology}\ }\textbf {\bibinfo {volume} {53}},\ \bibinfo {pages}
  {33-- 55} (\bibinfo {year} {1991}{\natexlab{a}})}\BibitemShut {NoStop}%
\bibitem [{\citenamefont {Kermack}\ and\ \citenamefont
  {McKendrick}(1991{\natexlab{b}})}]{kam_2}%
  \BibitemOpen
  \bibfield  {author} {\bibinfo {author} {\bibfnamefont {W.~O.}\ \bibnamefont
  {Kermack}}\ and\ \bibinfo {author} {\bibfnamefont {A.~G.}\ \bibnamefont
  {McKendrick}},\ }\bibfield  {title} {\enquote {\bibinfo {title}
  {Contributions to the mathematical theory of epidemics ii. the problem of
  endemicity},}\ }\href {\doibase 10.1007/BF02464424} {\bibfield  {journal}
  {\bibinfo  {journal} {Bulletin of Mathematical Biology}\ }\textbf {\bibinfo
  {volume} {53}},\ \bibinfo {pages} {57-- 87} (\bibinfo {year}
  {1991}{\natexlab{b}})}\BibitemShut {NoStop}%
\bibitem [{\citenamefont {Falco}\ \emph {et~al.}(2020)\citenamefont {Falco},
  \citenamefont {Cioppa}, \citenamefont {Scafuri},\ and\ \citenamefont
  {Tarantino}}]{rig}%
  \BibitemOpen
  \bibfield  {author} {\bibinfo {author} {\bibfnamefont {I.~D.}\ \bibnamefont
  {Falco}}, \bibinfo {author} {\bibfnamefont {A.~D.}\ \bibnamefont {Cioppa}},
  \bibinfo {author} {\bibfnamefont {U.}~\bibnamefont {Scafuri}}, \ and\
  \bibinfo {author} {\bibfnamefont {E.}~\bibnamefont {Tarantino}},\ }\href@noop
  {} {\enquote {\bibinfo {title} {Coronavirus covid-19 spreading in italy:
  optimizing an epidemiological model with dynamic social distancing through
  differential evolution},}\ } (\bibinfo {year} {2020}),\ \Eprint
  {http://arxiv.org/abs/2004.00553} {arXiv:2004.00553 [q-bio.PE]} \BibitemShut
  {NoStop}%
\end{thebibliography}%
\end{document}